\documentclass[10pt]{iopart}

\usepackage{iopams,setstack} 
\usepackage{cite} 
\usepackage{graphicx}
\usepackage{bm}
\usepackage{amssymb} 
\usepackage{hyperref} 
\usepackage{csvsimple-l3, booktabs} 
\usepackage[caption=false]{subfig} 
\usepackage[margin=30mm]{geometry}
\usepackage{xcolor} 
\hypersetup{colorlinks=true, anchorcolor=black} 

\begin{document}
	\title[]{Exploring the $ \boldsymbol{\Omega^-} $ spectrum in lattice QCD}
	\author{Liam Hockley, Waseem Kamleh,
		Derek Leinweber and Anthony Thomas}
	\address{ARC Special Research Centre for the Subatomic Structure of Matter (CSSM), Department of Physics, The University of Adelaide, SA, 5005,
		Australia.}
	\ead{liam.hockley@adelaide.edu.au}
	
	\begin{abstract}
		We present an exploratory lattice QCD analysis of the $ \Omega $-baryon spectrum. Using smeared three-quark operators in a correlation matrix analysis, we report masses for the ground, first and second excited states of the $ J^P = 1/2^\pm,\, 3/2^\pm $ spectra across a broad range in the light quark mass. We investigate the parity and spin quantum numbers for the states observed on the lattice, looking to reconcile these with the resonances encountered in experiment. We find that the $ \Omega^-(2012) $ as reported by the Particle Data Group corresponds to two overlapping resonances with $ J^P = 1/2^- $ and $ 3/2^- $. We also propose quantum number assignments for the higher energy resonances, and identify successive radial excitations within the spectra.
	\end{abstract}

	\noindent Keywords: lattice QCD, hadron spectroscopy, variational method, radial excitations

	\submitto{\jpg}


	\section{Introduction}\label{sec:OmegaIntro}
	One of the great triumphs of the quark model was the prediction of the existence of the $ \Omega^- $ baryon, completing the baryon decuplet formulated by Gell-Mann \cite{Gell-Mann:1961omu,Gell-Mann:1962yej} as the strangeness $ -3 $ state. Properties of the $ \Omega $ such as its spin being $ J=3/2 $ were first inferred from experiment in early measurements of $ K^-p $ interactions \cite{Aachen-Berlin-CERN-Innsbruck-London-Vienna:1977ojz} and $ \Omega^-\to \Lambda K^- $ decays \cite{Birmingham-CERN-Glasgow-MichiganState-Paris:1978nrw}. The spin assignment has since been verified by the BaBar collaboration \cite{BaBar:2006omx}, and more recently by the BESIII Collaboration through measurements of the angular distribution in the chain of decays $ \psi(3686) \to \Omega^- \overline{\Omega}^+ $, and $ \Omega^- \to K^- \Lambda $ and $ \overline{\Omega}^+ \to K^+ \overline{\Lambda} $ \cite{BESIII:2020lkm}.
	While the properties of the ground state $ \Omega^- $ resonance are well established, the situation for its excited states is very different. This is evidenced by \Tref{tab:OmegaPDG} where the question marks throughout the $ J^P $ column indicate unknown quantum numbers of the higher energy resonances. 
	
	\begin{table}[b]
		\caption{
			\label{tab:OmegaPDG}
			Table of observed resonances in the $ \Omega $ spectrum based on the PDG tables \cite{ParticleDataGroup:2022pth}, with masses $ m $, decay widths $ \Gamma $, and spin and parity quantum numbers $ J^P $. The classification scheme used is: 
			****   	Existence is certain, and properties are at least fairly explored.
			***   	Existence ranges from very likely to certain, but further confirmation is desirable and/or quantum numbers, branching fractions, etc. are not well determined.
			**   	Evidence of existence is only fair.
			*   	Evidence of existence is poor.
		}
		\begin{indented}
			\item[]\begin{tabular}{@{}lllll}
				\br 
				Resonance & $ m $ (MeV) & $ \Gamma $ (MeV) & $ J^P $ & Classification  \\
				\mr
				$ \Omega^-(1672) $ & $ 1672.5\pm \hphantom{5}0.3 $ & $ - $                 & $ 3/2^+ $ & **** \\
				$ \Omega^-(2012) $ & $ 2012.4\pm \hphantom{5}0.9 $ & $ 6.4^{+3.0}_{-2.6} $ & $ ?^- $ & *** \\
				$ \Omega^-(2250) $ & $ 2252\hphantom{.5}  \pm \hphantom{5}9   $ & $ 55 \pm 18 $         & $ ?^? $ & *** \\ 
				$ \Omega^-(2380) $ & $ 2380\hphantom{.5}  \pm17   $ & $ 26 \pm 23 $         & $ ?^? $ & ** \\
				$ \Omega^-(2470) $ & $ 2474\hphantom{.5}  \pm12   $ & $ 72 \pm 33 $         & $ ?^? $ & ** \\
				\br
			\end{tabular}
		\end{indented}
	\end{table}
	
	The second state in the spectrum is the $ \Omega^-(2012) $ which was first discovered in 2018 by the Belle collaboration \cite{Belle:2018mqs}. The PDG lists no spin quantum number, only indicating that it is expected to have odd parity. This is something we will confront in our analysis. Furthermore, the internal structure of the $ \Omega^-(2012) $ is an open question in terms of it being quark-model like or molecular \cite{Belle:2019zco,Ikeno:2020vqv,Belle:2022mrg, Lu:2022puv}. Needless to say, the nature of this resonance is far from resolved. Building a more complete understanding of the $ \Omega^-(2012) $ and the higher energy resonances in \Tref{tab:OmegaPDG}, of which even less is known, is an important endeavour in the current research landscape. New experiments at J-PARC \cite{Aoki:2021cqa, Sakuma:2022twx}, and potentially other next-generation colliders, will allow for much-needed experimental results to be obtained, and used to gain improved insight into the $ \Omega $ baryons. 
	
	Another piece of the puzzle lies in the regime of lattice QCD. Early studies of the spin-1/2 and spin-3/2 $ \Omega $ baryons on the lattice were conducted in studies such as that of Ref.~\cite{Engel:2013ig}; more recently the calculation performed in Ref.~\cite{Hudspith:2024kzk} offers high precision results for the masses of $ \Omega $ baryons incorporating $ \chi $PT analyses. In principle one would connect finite-volume energy levels from the lattice to experimental scattering data through L\"uscher's formalism \cite{Luscher:1985dn,Luscher:1986pf,Luscher:1990ux,Hansen:2012tf,Hansen:2014eka,Hansen:2019nir} or analogous techniques like Hamiltonian Effective Field Theory (HEFT) \cite{Hall:2013qba,Hall:2014gqa,Hall:2014uca,Liu:2015ktc,Liu:2016uzk,Liu:2016wxq,Wu:2017qve,Abell:2021awi,Abell:2023qgj,Leinweber:2024psf}. However, this is a challenging prospect since, as mentioned above, there is little in the way of experimental information concerning the $ \Omega $ baryons, and this includes the decay products needed for constraining HEFT or connecting with lattice QCD results through L\"uscher's method. Hence, in the following we look to lattice QCD as a first principles approach to gain insight into the $ \Omega $ spectrum.
	
	In this paper, we take the stance of studying the $ \Omega $ baryons as single-particle hadrons; we only use 3-quark operators in our analysis, following the same approach as used in Ref.~\cite{Hockley:2023yzn}. The resulting state masses are directly compared with the PDG values to see if they are in agreement with experiment. This is somewhat justified, at least for the lowest-lying $ \Omega $ baryons, since their decay widths are quite narrow, as evidenced by \Tref{tab:OmegaPDG}, for example only $ 6 $ MeV for the $ \Omega^-(2012) $. This indicates that couplings to multi-particle states may be small, and so for an exploratory study such as ours, we rely on sufficient Euclidean time evolution of lattice correlation functions to isolate states in the spectrum. 
	
	By performing parity and spin projections on our correlation functions, we aim to isolate distinct states in the spectrum and, where possible, use these to guide the identification of quantum numbers for the states in the spectrum. Additionally, we look to build on the work of Ref.~\cite{Hockley:2023yzn} by identifying radial excitations through the counting of nodes in the wave functions used to excite the states on the lattice. Our ability to identify radial excitations offers new insight unavailable in other studies of the $ \Omega $ spectrum.
	
	This paper is organised in the following way. \Sref{sec:SimDets} briefly outlines the PACS-CS gauge-field ensembles used in our calculations. The methods used for extracting effective masses and identifying radial excitations are explained in \Sref{sec:Method}, and we present the effective mass results and wave function node identification in Section~\ref{sec:OmegaResults}. In Section~\ref{sec:Remarks} we compare the masses obtained on the lattice with those listed in the PDG, focusing on quantum number assignment for the resonances. We present our conclusions in Section~\ref{sec:Conclusion}.
	
	\section{Simulation Details} \label{sec:SimDets}
	Our calculations are performed on the PACS-CS Collaboration's gauge-field ensembles \cite{PACS-CS:2008bkb} made available via the ILDG \cite{Beckett:2009cb}. The same ensembles were used in Ref.~\cite{Hockley:2023yzn}. These configurations make use of the Iwasaki gauge action \cite{Iwasaki:1985we} and an $ \mathcal{O}(a) $-improved Wilson quark action with Clover coefficient $ c_\mathrm{SW} = 1.715 $, in full $ 2+1 $ flavour dynamical QCD. The simulations use a gauge field strength parameter of $ \beta=1.90 $ on a $ 32^3\times 64 $ lattice. 
	
	The masses of the up and down quarks are taken to be degenerate, having hopping parameter $ \kappa_{ud} $. Meanwhile, we hold the strange quark mass fixed at the hopping parameter value $ \kappa_s = 0.13665 $, chosen such that kaons have physical masses \cite{Menadue:2012kc}. Using the correct $ \kappa_s $ value is particularly important here since the $ \Omega $ baryons are comprised of three strange valence quarks. Note that since we use the PACS-CS ensembles to begin with, the strange quark masses in the sea are slightly mistuned, using the value $ \kappa_s = 0.13640 $. A future lattice calculation should use the corrected strange quark mass for \textit{both} the sea and valence quarks. The characteristics of each ensemble are given in \Tref{tab:LatSimParams}. 
	
	\begin{table}
		\caption{
			\label{tab:LatSimParams}
			Table of parameters for the lattice ensembles used in our calculations, characterised by the hopping parameter $ \kappa $. The configurations are calculated on lattices with spacings $ a $ and spatial lengths $ L $, with ensembles of size $ N_\mathrm{cfgs} $. The gauge fields are sampled $ N_\mathrm{src} $ times and averaged over the number of samples in constructing our correlation functions.
		}
		\begin{indented}
			\item[]\begin{tabular}{@{}llllll}
				\br 
				$ \kappa $ & $ m_\pi $ (MeV) & $ a $ (fm) & $ L $ (fm) & $ N_\mathrm{cfgs} $ & $ N_\mathrm{src} $ \\
				\mr
				0.13781 & 169 & 0.0933(13) & 2.9856(416) & 198 & 64 \\
				0.13770 & 280 & 0.0951(13) & 3.0432(416) & 400 & 32 \\
				0.13754 & 391 & 0.0961(13) & 3.0752(416) & 449 & 32 \\
				0.13727 & 515 & 0.1009(15) & 3.2288(480) & 397 & 32 \\
				0.13700 & 623 & 0.1023(15) & 3.2736(480) & 399 & 16 \\
				\br
			\end{tabular}
		\end{indented}
	\end{table}	
	
	\section{Methodology} \label{sec:Method}
	
	\subsection{Hadron Spectroscopy} \label{subsec:HadSpec}
	In this work, lattice QCD provides a means for us to calculate masses for the $ \Omega $-baryon spectrum using established tools within the field of hadron spectroscopy. These follow the same approach used in Ref.~\cite{Hockley:2023yzn} and we briefly summarise them here.
	
	The energies in the spectrum of a particular baryon species can be derived from 2-point correlation functions.	At the hadronic level, a general 2-point correlation function includes Fourier projection to states of definite momentum $ \boldsymbol{p} $, and is defined as
	\begin{equation}
		\mathcal{G}^{ij}_{\mu\nu}(t,\boldsymbol{p}) = \sum_{\boldsymbol{x}} e^{-i\boldsymbol{p}\cdot \boldsymbol{x}} 
		\langle {0} | {T\{ \chi^i_\mu(x) \bar{\chi}^j_\nu(0) \}} | {0} \rangle
		\,. \label{eq:2point}
	\end{equation}
	The indices $ i, j $ refer to choices of interpolating field $ \chi $, with Greek indices $ \mu,\nu $ labelling the Lorentz degrees of freedom. $ T $ indicates time ordering of the operators. The operator $ \bar{\chi}^j_\nu(0) $ acts on the vacuum $ | {0} \rangle $ to create states at the space-time point $ (0, \boldsymbol{0}) $. These states then propagate through Euclidean time $ t $ before being annihilated at a new space-time point $ x = (t,\boldsymbol{x}) $ by the operator $ \chi^i_\mu(x) $.	One can then show \cite{Hockley:2023yzn} that, by introducing a complete set of states, the correlation function may be written as
	\begin{eqnarray}
		\mathcal{G}^{ij}_{\mu\nu}(t,\boldsymbol{p}) 
		= \sum_{B,s} &
		e^{-E_Bt} 
		\langle {0} | {\chi^i_\mu(0)} | {B,\boldsymbol{p},s} \rangle 
		\langle {B,\boldsymbol{p},s} | {\bar{\chi}^j_\nu(0)} | {0} \rangle
		\,, \label{eq:correlator}
	\end{eqnarray}
	where $ B $ is the baryon species label and $ s $ is the spin quantum number.
	
	In this work we take the well known interpolating field for the $ \Delta^{++} $ given as 
	\begin{equation}
		\chi^{\Delta^{++}}_\mu(x) = \epsilon^{abc}\, \big(u^{Ta}(x)\, C \gamma_\mu\, u^b(x)\big)\, u^c(x)\,,
	\end{equation}
	and replace each $ u $ quark with an $ s $ quark, yielding
	\begin{equation}
		\chi^{\Omega^-}_\mu(x) = \epsilon^{abc}\, \big(s^{Ta}(x)\, C \gamma_\mu\, s^b(x)\big)\, s^c(x)\,. \label{eq:omega_interp}
	\end{equation}
	Here $ u(x) $ and $ s(x) $ represent the Dirac spinor for a single $ u $ or $ s $ quark at spacetime position $ x $ carrying a colour index $ a,b,c $. In the Dirac representation of the $ \gamma $-matrices which we use here, the charge conjugation matrix takes the form $ C = i\gamma_2\gamma_0 $. $ \epsilon $ is the Levi-Cevita antisymmetric tensor. Moving forward, interpolating fields are only used to describe $ \Omega $ baryons and so we drop the $ \Omega^- $ superscript for ease of notation.
	
	Given that each of the three quarks in $ sss $ is a spin-$ 1/2 $ particle, the addition of their angular momenta yields states with either spin-$ 1/2 $ or spin-3/2. To distinguish between the various spin states, we need to apply spin-projection operators to the correlation functions. This is discussed further in \Sref{subsec:SpinProj}. 
	
	We note that given the parity transformation property of the strange-quark fermion spinors
	\begin{equation}
		s(x) \to \gamma_0\, s(\tilde{x})\,,
	\end{equation}
	where $ \tilde{x} \equiv (t,-\boldsymbol{x}) $, the spatial components of the $ \Omega^{-} $ interpolating field transform as
	\begin{equation}
		\chi_i(x) \to +\, \gamma_0\, \chi_i(\tilde{x})\,, \quad i=1,2,3\,, \label{eq:parity_transform}
	\end{equation}
	while the temporal component transforms as
	\begin{equation}
		\chi_0(x) \to -\, \gamma_0\, \chi_0(\tilde{x})\,. \label{eq:parity_transform2}
	\end{equation}
	
	Together, Equations \eref{eq:parity_transform} and \eref{eq:parity_transform2} describe the parity transformation property of a pseudovector times a spinor, a property shared by the Rarita-Schwinger spin-vector $ u_\mu(x) $. Thus, for our case of interest we have the following available matrix elements, with the associated decomposition into Dirac spinors $ u(\boldsymbol{p},s) $ for the spin-1/2 components, and Rarita-Schwinger spinors $ u_\mu(\boldsymbol{p},s) $ for the spin-3/2 components \cite{Belyaev:1982sa}:
	
	\begin{eqnarray}
		& \langle {0} | {\chi^i_\mu(0)} | {\Omega^{3/2^+}(\boldsymbol{p},s)} \rangle = \lambda_{3/2^+} \sqrt{\frac{M_{3/2^+}}{E_{3/2^+}}}\, u_\mu (\boldsymbol{p},s)\,, \label{eq:matprod1} \\
		& \langle {0} | {\chi^i_\mu(0)} | {\Omega^{3/2^-}(\boldsymbol{p},s)} \rangle = \lambda_{3/2^-} \sqrt{\frac{M_{3/2^-}}{E_{3/2^-}}}\, \gamma_5 \, u_\mu (\boldsymbol{p},s)\,, \label{eq:matprod2} \\
		& \langle {0} | {\chi^i_\mu(0)} | {\Omega^{1/2^+}(\boldsymbol{p},s)} \rangle = (\alpha_{1/2^+}p_\mu + \beta_{1/2^+}\gamma_\mu)
		\sqrt{\frac{M_{1/2^+}}{E_{1/2^+}}} \, \gamma_5 \, u (\boldsymbol{p},s)\,, \label{eq:matprod3} \\
		& \langle {0} | {\chi^i_\mu(0)} | {\Omega^{1/2^-}(\boldsymbol{p},s)} \rangle = (\alpha_{1/2^-}p_\mu + \beta_{1/2^-}\gamma_\mu) 
		\sqrt{\frac{M_{1/2^-}}{E_{1/2^-}}}\, u(\boldsymbol{p},s)\,. \label{eq:matprod4} 
	\end{eqnarray}
	
	Here, the $ \lambda $, $ \alpha $ and $ \beta $ factors are coupling strengths and $ M $ is the mass of the state with relativistic energy $ E $. \Eref{eq:matprod1} reflects the fact that both the Rarita-Schwinger spinor and the interpolating field transform under parity as a pseudovector. The square root factor accounts for the differing normalisations of the Rarita-Schwinger spinor and the full baryon wave function. $ \gamma_5 $ matrices are placed throughout to produce matrix elements with the correct parity transformation properties as described in Ref.~\cite{Hockley:2023yzn}.
	
	Including the corresponding expressions for the $ \bar{\chi} $ matrix elements from \Eref{eq:correlator}, one can use the Dirac spinor spin-sum relation \cite{Itzykson:1980rh}
	\begin{equation}
		\sum_{s = -1/2}^{1/2} u(\boldsymbol{p},s) \bar{u}(\boldsymbol{p},s) = \frac{\gamma \cdot p + M_B}{2M_B}\,, \label{eq:diracspinsum}
	\end{equation}
	and the analogous Rarita-Schwinger spin-sum relation \cite{Zanotti:2003fx}
	
	\begin{equation}
		\sum_{s = -3/2}^{3/2} u_\mu(\boldsymbol{p},s) \bar{u}_\nu(\boldsymbol{p},s) 
		= 
		-\frac{\gamma \cdot p + M_B}{2M_B} \bigg( g_{\mu\nu} - \frac{1}{3}\gamma_\mu \gamma_\nu - \frac{2p_\mu p_\nu}{3M_B^2} + \frac{p_\mu \gamma_\nu - p_\nu \gamma_\mu}{3M_B} \bigg) 
		\,, \label{rsspinsum}
	\end{equation}
	to compute the sum over $ s $ and write the contributions from each of the required products of matrix elements. At $ \boldsymbol{p} = 0 $, so that $ E_{B}~=~M_{B} $, and with $ \mu = \nu = n = 1,2,3 $, we have
	\begin{eqnarray}
		& 
		\sum_s\, \langle {0} | {\chi^i_n(0)} | {\Omega^{3/2^+}(\boldsymbol{0},s)} \rangle 
		\langle {\Omega^{3/2^+}(\boldsymbol{0},s)} | {\bar{\chi}^j_n(0)} | {0} \rangle 
		= \lambda^i_{3/2^+}\, \overline{\lambda}^{j}_{3/2^+}\, \frac{2}{3} \bigg( \frac{\gamma_0 + \mathbb{I}}{2} \bigg)\,, \\
		&
		\sum_s\, \langle {0} | {\chi^i_n(0)} | {\Omega^{3/2^-}(\boldsymbol{0},s)} \rangle
		\langle {\Omega^{3/2^-}(\boldsymbol{0},s)} | {\bar{\chi}^j_n(0)} | {0} \rangle 
		= \lambda^i_{3/2^-}\, \overline{\lambda}^{j}_{3/2^-}\, \frac{2}{3} \bigg( \frac{\gamma_0 - \mathbb{I}}{2} \bigg)\,, \\
		&
		\sum_s\, \langle {0} | {\chi^i_n(0)} | {\Omega^{1/2^+}(\boldsymbol{0},s)} \rangle
		\langle {\Omega^{1/2^+}(\boldsymbol{0},s)} | {\bar{\chi}^j_n(0)} | {0} \rangle 
		= \lambda^i_{1/2^+}\, \overline{\lambda}^{j}_{1/2^+}\, \bigg( \frac{\gamma_0 + \mathbb{I}}{2} \bigg)\,, \\
		&
		\sum_s\, \langle {0} | {\chi^i_n(0)} | {\Omega^{1/2^-}(\boldsymbol{0},s)} \rangle
		\langle {\Omega^{1/2^-}(\boldsymbol{0},s)} | {\bar{\chi}^j_n(0)} | {0} \rangle 
		= \lambda^i_{1/2^-}\, \overline{\lambda}^{j}_{1/2^-}\, \bigg( \frac{\gamma_0 - \mathbb{I}}{2} \bigg)\,,
	\end{eqnarray}
	where we've relabelled $ \beta_{1/2^\pm} \to \lambda_{1/2^\pm} $ for simplicity. The notation $ \overline{\lambda} $ is used to denote the overlap of the interpolator at the source.
	
	Hence we can see that the even-parity states exist in the $ (1,1) $ and $ (2,2) $ Dirac components of the correlation function, while the odd-parity states live in the $ (3,3) $ and $ (4,4) $ Dirac components. These are then readily accessed by applying the appropriate projection operators	
	\begin{equation}
		\Gamma^\pm = \frac{1}{2} (\gamma_0 \pm \mathbb{I})\,, \label{eq:ParProj}
	\end{equation}
	where $ \mathbb{I} $ is the identity operator. In particular, we note that the parity information of our correlation functions is governed by the upper and lower components of the spinors describing the baryon states. This will be of importance in Section~\ref{subsec:RadExc} in discussing the identification of radial excitations for odd-parity states of the $ \Omega $.
	
	One can then define the parity-projected correlator as the trace over Dirac indices of the sum over the spatial Lorentz indices $ \mu = \nu = n $ \cite{Leinweber:1992hy}
	\begin{equation}
		[G^\pm(t,\boldsymbol{0})]^{ij} \equiv \Tr \bigg[\Gamma^\pm \sum_{n=1}^3 \mathcal{G}^{ij}_{nn}(t,\boldsymbol{0}) \bigg]
		\,.  \label{eq:trace}
	\end{equation}
	
	Computing the right-hand-side yields a series of decaying exponentials governed by the mass of the baryon states
	\begin{equation}
		[G^\pm(t,\boldsymbol{0})]^{ij} = \sum_{B^\pm} \lambda_{B^\pm}^{i}\, \overline{\lambda}_{B^\pm}^{j}\, e^{-M_{B^\pm}t}\,, \label{eq:expl_series}
	\end{equation}
	where $ \lambda^{i}_{B^\pm} $ and $ \overline{\lambda}^j_{B^\pm} $ are coupling strengths between the interpolating fields $ \chi^i_\mu $ and $ \overline{\chi}^j_\nu $ and the parity projected baryon states $ B^{\pm} $, up to some overall constant factors. We note that these coupling strengths can be taken to be real by considering both the original gauge-field links and their complex conjugates, weighted equally in the ensemble average \cite{Mahbub:2013ala,Leinweber:1992hy}. 
	
	The parity projected correlation function still contains various spin states (such as spin-1/2 and spin-3/2 in our case) as well as a tower of excited states. In order to extract, say, the ground state mass of a particular baryon state, one takes the long-time limit in which all the excited states have decayed off. Explicitly,
	\begin{equation}
		[G^\pm(t,\boldsymbol{0})]^{ij} \overset{t\to\infty}{=} \lambda_{0^\pm}^{i}\, \overline{\lambda}_{0^\pm}^{j}\, e^{-M_{0^\pm}t}\,, \label{eq:groundcorr}
	\end{equation}
	where the $ \lambda_{0^\pm}^{i} $ and $\overline{\lambda}_{0^\pm}^{j} $ are couplings of baryon interpolators at the source and sink to the lowest-lying state, having even ($ + $) or odd ($ - $) parity. In order to isolate spin states one can employ further projection techniques as discussed in the following section. A discussion of how one can access excited energy states is presented in Section~\ref{subsec:VarMeth}.
	
	\subsection{Spin Projection} \label{subsec:SpinProj} 
	As highlighted earlier, our choice of a 3-quark operator to descibe the $ \Omega^{-} $ baryons couples to both spin-1/2 and spin-3/2 states. In order to project these states, we employ projection operators as given in Refs.~\cite{Zanotti:2003fx,Benmerrouche:1989uc}
	\begin{equation}
		P^{3/2}_{\mu\nu} (p) = g_{\mu\nu} - \frac{1}{3}\gamma_\mu\gamma_\nu - \frac{1}{3p^2}(\gamma\cdot p\ \gamma_\mu p_\nu + p_\mu \gamma_\nu\ \gamma\cdot p)\,, \label{eq:proj32}
	\end{equation}
	\begin{equation}
		P^{1/2}_{\mu\nu} (p) = g_{\mu\nu} - P^{3/2}_{\mu\nu}(p)\,. \label{eq:proj12}
	\end{equation}
	
	Although \Eref{eq:proj32} looks somewhat cumbersome, we can do a few things to simplify these operators. First, in our lattice calculations the energy eigenstates are on shell, so we have
	\begin{equation}
		p = (E,\boldsymbol{p}) = (\sqrt{\boldsymbol{p}^2 + m^2},\boldsymbol{p})\,.
	\end{equation}
	
	We consider the particles at rest, $ \boldsymbol{p} = 0 $, simplifying our spin-3/2 projection operator to
	\begin{eqnarray}
		P^{3/2}_{\mu\nu} (\boldsymbol{p}=0)
		&= g_{\mu\nu} - \frac{1}{3}\gamma_\mu\gamma_\nu - \frac{1}{3}(\gamma_0 \gamma_\mu g_{\nu 0} + g_{\mu 0} \gamma_\nu \gamma_0)\,. \label{eq:reduced}
	\end{eqnarray}
	
	One can then show by using the properties of the $ \gamma $-matrices and the metric, that the elements of the projector obey
	\begin{eqnarray}
		&P^{3/2}_{0 0} (\boldsymbol{p}=0) = 
		P^{3/2}_{0 n} (\boldsymbol{p}=0) = 
		P^{3/2}_{m 0} (\boldsymbol{p}=0) = 0\,, \label{eq:res0} \\
		&P^{3/2}_{m n} (\boldsymbol{p}=0) = g_{m n} - \frac{1}{3}\gamma_m\gamma_n \,, \label{eq:res1}
	\end{eqnarray}
	where $ m,\, n $ are spatial Lorentz indices.
	
	We get the corresponding forms for the spin-1/2 projection operator by making use of \Eref{eq:proj12}, \eref{eq:res0} and \eref{eq:res1} to obtain
	\begin{eqnarray}
		&P^{1/2}_{0 0} (\boldsymbol{p}=0) = \mathbb{I}\,, \\
		&P^{1/2}_{0 n} (\boldsymbol{p}=0) =
		P^{1/2}_{m 0} (\boldsymbol{p}=0) = 0\,, \\
		&P^{1/2}_{m n} (\boldsymbol{p}=0) = \frac{1}{3}\gamma_m\gamma_n\,. \label{eq:res2}
	\end{eqnarray}
	
	With the spin projection operators in hand, a spin-$ s $ projected correlation function is then given by
	\begin{equation}
		[\mathcal{G}^s_{\mu\nu}]^{ij} = \sum_{\sigma,\lambda = 1}^{4} \mathcal{G}^{ij}_{\mu\sigma} g^{\sigma\lambda} P^s_{\lambda\nu}\,.
	\end{equation}
	
	This spin projection is performed prior to the parity projection and trace in \Eref{eq:trace}. For completeness, a spin-$ s $ and parity-projected correlation function is written as
	\begin{equation}
		[G^{s^\pm}(t,\boldsymbol{0})]^{ij}  \equiv \Tr \bigg(\Gamma^\pm \sum_{n=1}^3  [\mathcal{G}^s_{nn}(t,\boldsymbol{0})]^{ij} \bigg)
		\,.
	\end{equation}
	
	\subsection{Source and Sink Smearing}\label{subsec:SourSinkSmear}
\begin{figure}[]
		\centering
		\includegraphics[width=0.4\linewidth]{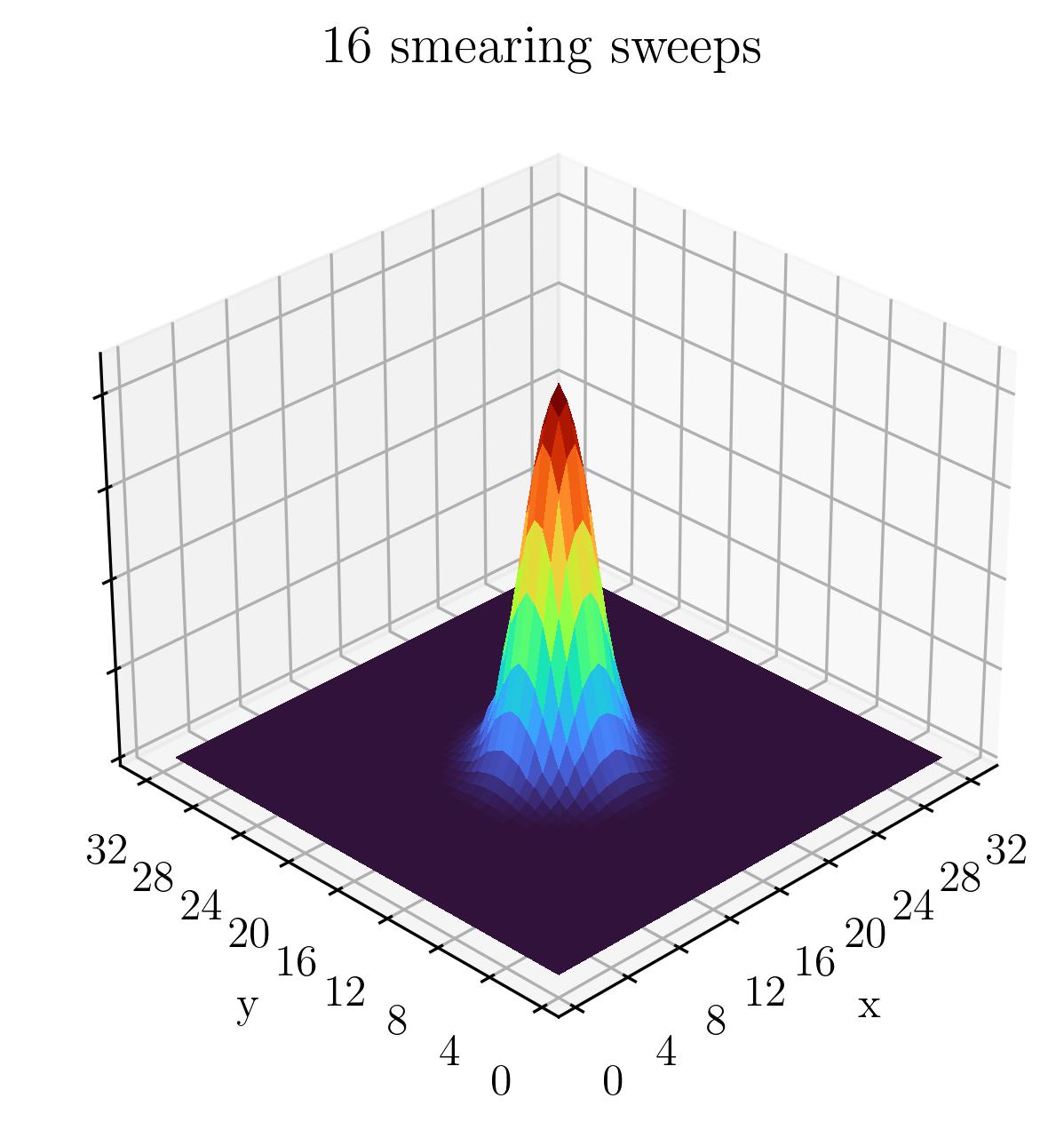}\hfil
		\includegraphics[width=0.4\linewidth]{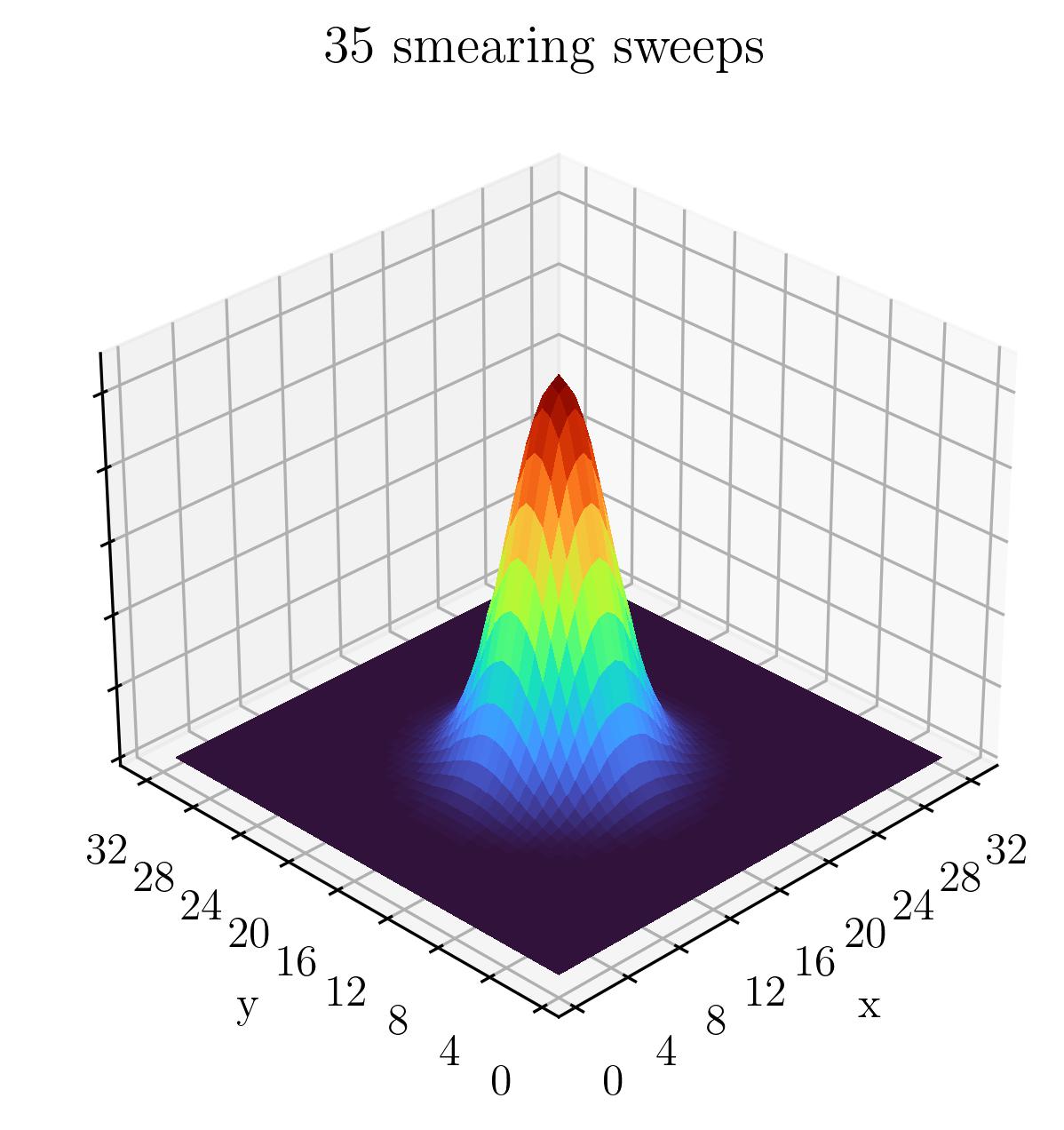}
		
		\includegraphics[width=0.4\linewidth]{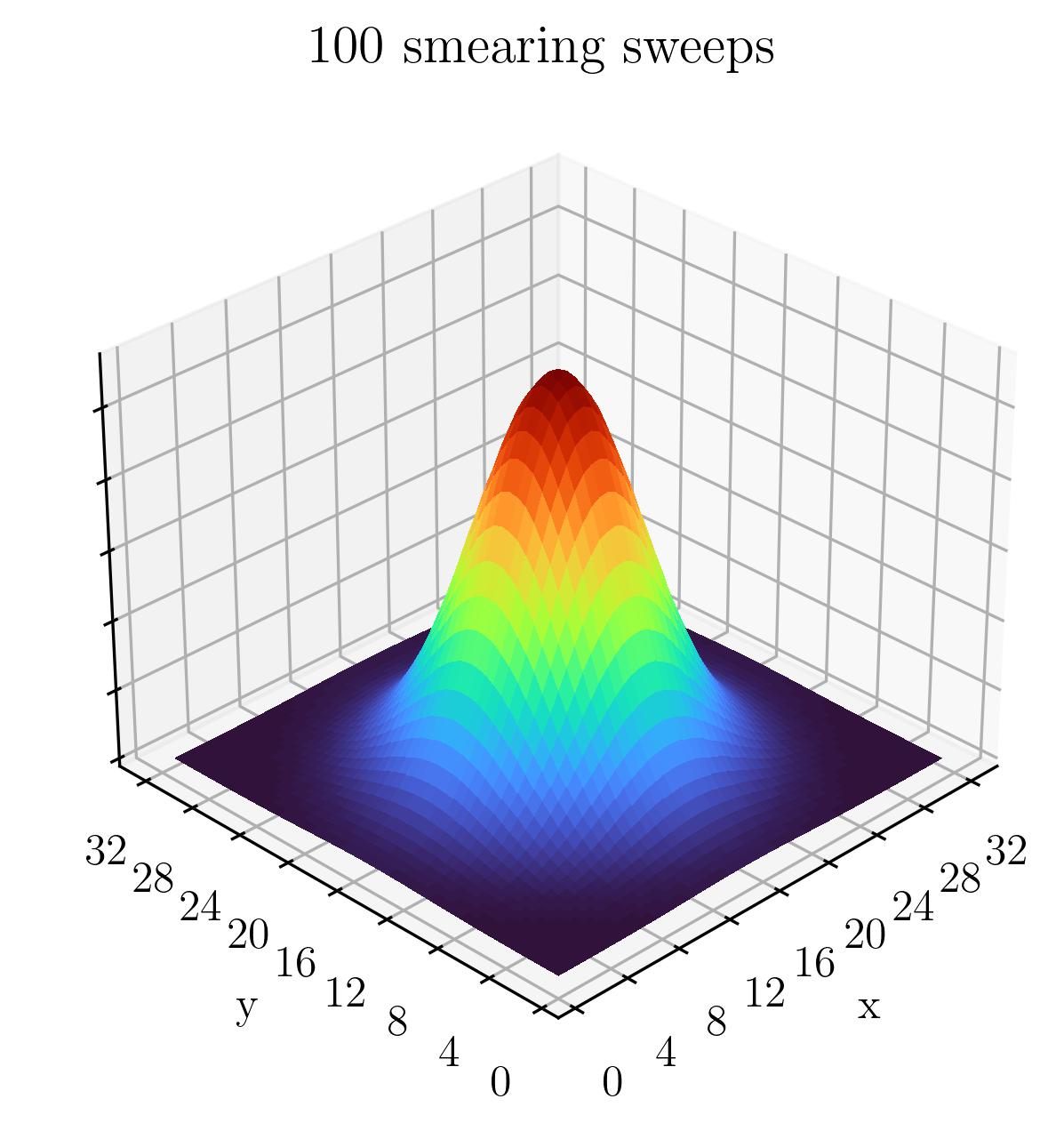}\hfil
		\includegraphics[width=0.4\linewidth]{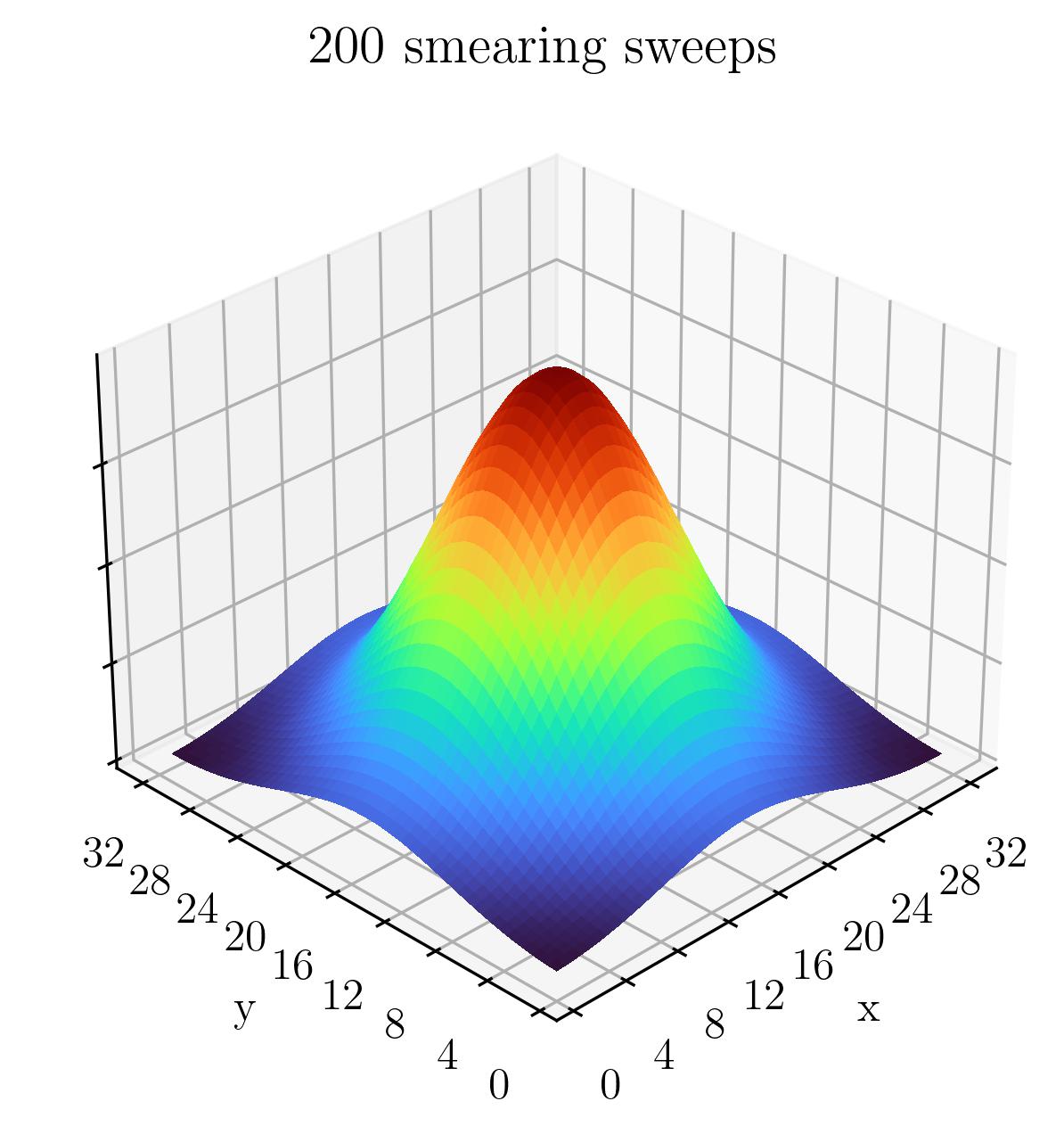}
		\caption{Plots of point sources after applying the smearing operation 16, 35, 100 and 200  times. The height of the peak is kept fixed in the visualisation to illustrate the broadening of the quark distribution.}
		\label{fig:SmearGauss}
	\end{figure}

	In order to improve the overlap of the interpolating fields with the states of interest, we apply Gaussian smearing to the spatial components of the quark operators. The general procedure is to take some fermion field $ \psi_0(t, \boldsymbol{x}) $ and iteratively apply a smearing function $ F(\boldsymbol{x},\boldsymbol{x}') $. Explicitly, this takes the form
	\begin{equation}
		\psi_i(t,\boldsymbol{x}) = \sum_{\boldsymbol{x}'}\, F(\boldsymbol{x},\boldsymbol{x}')\, \psi_{i-1}(t,\boldsymbol{x}')
		\,,
	\end{equation}
	where the smearing function is
	\begin{eqnarray}
		F(\boldsymbol{x},\boldsymbol{x}') &= (1-\alpha)\,\delta_{x,x'} + \frac{\alpha}{6} \sum_{\mu = 1}^{3} \big[U_\mu(x)\,\delta_{x',x+\hat{\mu}} + U^\dagger_\mu(x - \hat{\mu})\,\delta_{x',x-\hat{\mu}}\big]
		\,.
	\end{eqnarray}
	We take the smearing parameter to be $ \alpha=0.7 $ in our calculations. The use of repeated applications of the smearing function controls the width of our source by gradually smearing out an initial point source. This is visualised in \Fref{fig:SmearGauss} where, for 16, 35, 100 and 200 smearing iterations or \textit{sweeps}, the amplitude of the distribution for $ U_\mu(x) = \mathbb{I} $ is plotted on the $ x-y $ plane with the third spatial dimension fixed at the centre of the source. 
	
	\subsection{Variational Method}\label{subsec:VarMeth} 
	Recall that in \Sref{subsec:HadSpec} we showed how one can readily obtain the mass of baryon ground states by simply taking the leading order contribution to the correlation function as in \Eref{eq:groundcorr}. In order to isolate states of higher energies, a more nuanced approach is required, since these states are at sub-leading order in the exponential series. We make use of the well-known variational analysis method \cite{Michael:1985ne, Mahbub:2013ala} in order to extract the ground state mass as well as the excited state masses.
	
	To extract $ N $ states, we require $ N $ interpolating fields (one for each state). We generate our interpolating fields by using the same base interpolator for the $ \Omega^- $ baryon, obtained from \Eref{eq:omega_interp} and applying different levels of smearing consisting of 16, 35, 100 or 200 sweeps of smearing, as discussed in Section~\ref{subsec:SourSinkSmear}. 
	
	
	Leaving all parity and spin labels implicit (in other words assuming the correlation functions have been parity and spin projected), the correlation matrix is written as
	\begin{equation}
		G^{ij}(t) = \sum_{\alpha=0}^{N-1}\, \lambda_{\alpha}^i \, \overline{\lambda}_{\alpha}^j \, e^{-m_{\alpha} t}\,. \label{corr3}
	\end{equation}
	Here the $ \lambda_\alpha^i $ and $ \overline{\lambda}_\alpha^j $ are essentially the same as seen before in Section~\ref{subsec:HadSpec}, though we now use the $ \alpha $ index to distinguish between energy states. In other words, they are couplings of the smeared interpolators $ \chi^i_\mu $ and $ \overline{\chi}^j_\nu $ at the sink and source, respectively, to the various energy eigenstates $ \alpha = 0, \dots , N-1 $. $ m_{\alpha} $ is the mass of the energy eigenstate $ \alpha $. 
	
	From here, we now aim to construct linear combinations of our smeared interpolating fields to cleanly isolate the $ N $ states in the baryon spectrum. Labelling these baryon states $ | {B_\alpha} \rangle $, we thus wish to construct the superpositions
	\begin{equation}
		(\overline{\phi}_\alpha)_\mu = \sum_{i = 1}^{N} u_\alpha^i\, \overline{\chi}^i_\mu 
		\quad \mathrm{and} \quad 
		(\phi_\alpha)_\nu = \sum_{i = 1}^{N} v_\alpha^i\, \chi^i_\nu
		\,, \label{phibar}
	\end{equation}
	such that
	\begin{eqnarray}
		\langle {B_\beta, \boldsymbol{p},s} | (\overline{\phi}_\alpha)_\mu | {0} \rangle 
		= 
		\delta_{\alpha\beta}\, \overline{z}_\alpha\, \overline{u}_\mu(\alpha,\boldsymbol{p},s) 
		\\ 
		\langle {0} | {(\phi_\alpha)_\nu} | {B_\beta, \boldsymbol{p},s} \rangle 
		= 
		\delta_{\alpha\beta}\, z_\alpha\, u_\nu(\alpha,\boldsymbol{p},s)
		\,, \label{phi}
	\end{eqnarray}
	where $ u_\mu(\alpha, \boldsymbol{p},s) $ is a Rarita-Schwinger spin vector. Here, the $ z_\alpha $ and $ \overline{z}_\alpha $ are the couplings of the superpositions $ \phi_\alpha $ and $ \overline{\phi}_\alpha $ to the state $ | {B_\alpha} \rangle $. The $ u_\alpha^i $ and $ v_\alpha^i $ are simply the weights for the basis of smeared interpolating fields.
	
	At this point, we construct an eigenvalue problem to solve for both $ \boldsymbol{u}_\alpha $ and $ \boldsymbol{v}_\alpha $. Noting that since $ G^{ij}(t) $ is real and symmetric, $ G^{ij}(t) = G^{ji}(t) $, in the ensemble average we introduce an improved unbiased estimator of the correlation matrix $ [G^{ij}(t) + G^{ji}(t)]/2 $. This provides us with a correlation matrix which is symmetric, so we can simultaneously compute $ \boldsymbol{u}_\alpha $ and $ \boldsymbol{v}_\alpha $ as discussed below.
	
	Multiplying the correlation matrix on the right by $ u_\alpha^j $ we obtain
	\begin{equation}
		G^{ij}(t)\, u_\alpha^j = \lambda_\alpha^i \, \overline{z}_\alpha\, e^{-m_\alpha t}\,.
	\end{equation}
		
	Then, since the exponential is the only time-dependent part of the correlation function, we can form a  recurrence relation at some time after source insertion by introducing the variational parameters $ t_0 $ and $ \Delta t $:
	\begin{equation}
		G^{ij}(t_0 + \Delta t)\, u_\alpha^j = e^{-m_\alpha \Delta t}\, G^{ij}(t_0)\, u_\alpha^j\,.
	\end{equation}
	
	Then, multiplying on the left by the inverse $ [G^{ij}(t_0)]^{-1} $ and suppressing the indices $ i $ and $ j $ gives 
	\begin{equation}
		[G(t_0)^{-1} G(t_0 + \Delta t)]\, \boldsymbol{u}_\alpha = e^{-m_\alpha \Delta t}\,\boldsymbol{u}_\alpha\,, \label{genu}
	\end{equation}
	which we recognise as an eigenvalue equation for the vector in interpolator space $ \boldsymbol{u}_\alpha $.
	
	Similarly, premultiplying the correlation matrix by $ v^\alpha_i $ we get
	\begin{equation}
		v_\alpha^i G^{ij}\,(t_0 + \Delta t) = e^{-m_\alpha \Delta t}\, v_\alpha^i \, G^{ij}(t_0)
	\end{equation}
	from which we arrive at our second eigenvalue equation (this time for $ \boldsymbol{v}_\alpha $):
	\begin{equation}
		\boldsymbol{v}_\alpha\, G(t_0 + \Delta t)\,[G(t_0)]^{-1} = e^{-m_\alpha \Delta t}\, \boldsymbol{v}_\alpha\,. \label{genv}
	\end{equation}
	
	Both \Eref{genu} and \eref{genv} need to be solved simultaneously for each given pair of variational parameters $ t_0 $ and $ \Delta t $, and we do so using a generalised eigenvalue problem (GEVP) solver. Solving for these eigenvectors automatically gives us the weights for the superpositions of interpolating fields, by construction. 
	
	Finally, the eigenstate-projected correlation function is then taken to be
	\begin{equation}
		G_\alpha(t) \equiv v_\alpha^i \, G^{ij}(t)\, u_\alpha^j\,.
	\end{equation}
	The eigenvectors are essentially used to isolate particular states in the baryon spectrum, exactly as we set out to do. Replacing the spin-index $ s $ and parity label $ \pm $ we have the full parity-, spin- and energy-eigenstate-projected correlation function
	\begin{equation}
		G^{s^\pm}_\alpha(t) = v_\alpha^i \, [G^{s^\pm}_\alpha(t)]^{ij}\, u_\alpha^j\,.
	\end{equation}

	We then construct the effective mass function
	\begin{equation}
		[M_\alpha^{s^\pm}(t)] = \frac{1}{\delta t} \ln\bigg(\frac{G^{s^\pm}_\alpha(t,\boldsymbol{0})}{G^{s^\pm}_\alpha(t+\delta t,\boldsymbol{0})}\bigg)
		\,. \label{effmass}
	\end{equation}
	$ \delta t $ is typically taken to be small and is set independently of the variational parameters. We take $ \delta t = 2 $ in our calculations.
	
	It is also worth noting that the eigenvectors $ \boldsymbol{u}_\alpha $ and $ \boldsymbol{v}_\alpha $ are equal since $ G^{ij}(t) $ is a real symmetric matrix. From here on, we will refer to the $ \boldsymbol{u}_\alpha $ vector, for simplicity.
	
	The effective mass defined in \Eref{effmass} can be computed for various discrete values of $ t $ and then plotted as a function of Euclidean time. As usual, one looks for time intervals over which the effective mass plot plateaus, indicating that all contamination from unwanted excited states has decayed away in the exponential series. We then perform a covariance matrix analysis of the $ \chi^2 $ per degree of freedom \cite{Mahbub:2013bba} to determine the most suitable time intervals to fit when obtaining our final masses. The details of this procedure for calculating our fit uncertainties are given in Section 5 of Ref.~\cite{Mahbub:2013bba}.
	
	With a basis of 4 smeared interpolating fields, we have access to 4 energy eigenstates in principle. However, while the third excited state is available to us, it is susceptible to excited state contamination. In some cases the signal-to-noise ratio degrades too rapidly for us to reliably report results, and these are omitted. Hence we only report the masses of the ground, first and second excited states.
	
	As a final note, the principles highlighted within this section are fully realised only when one includes a complete set of interpolating fields effective at isolating all the states within the spectrum. In principle this needs to include multi-particle scattering states as highlighted in, for example, Ref.~\cite{Hansen:2019nir}. However, our aim is simply to apply the same techniques as used in Ref.~\cite{Hockley:2023yzn}, and give some new insight into the spectrum of $ \Omega $ baryons. Our formalism is suited to exciting these single-particle states and in particular identifying their radial node structure. As such, Euclidean time evolution is important in suppressing contamination from nearby states. In fitting our effective mass to a plateau, we enhance single-state dominance by monitoring the $ \chi^2 $ per degree of freedom ($ \chi^2/\mathrm{dof} $). We enforce an upper limit of $ \chi^2/\mathrm{dof} \leq 1.2 $ for our fits.
	
	\subsection{Identifying Radial Excitations} \label{subsec:RadExc}
	A key feature of this study is the ability to identify radial excitations within the spectrum, and following Ref.~\cite{Hockley:2023yzn}, this naturally falls out of the variational method outlined in Section~\ref{subsec:VarMeth}. Recall that the variational method relies on constructing an improved estimate of the interpolating field from a superposition of smeared interpolating fields,
	\begin{equation}
		\phi_\alpha = \sum_{i}\, u_\alpha^i \chi^i\,,
	\end{equation}
	where $ i = 16,35,100,200 $ denotes the number of smearing sweeps. Essentially, solving the GEVP tells us how to construct the wave function for a given energy eigenstate from a basis of Gaussians; an important benefit of this is that such a superposition can give rise to nodes in the wave function. While a distribution with no nodes indicates a $ 1s $ state, successive radial excitations of this state can be identified by the presence of one or more nodes in the wave function.
	
	Consider, for example, the case of a narrow Gaussian of positive signature, superposed with a broader Gaussian of negative signature. As we consider points further away from the centre of the distribution, the positive Gaussian will diminish and the negative Gaussian will begin to dominate the superposition. As a result, there will be some radius at which the superposition crosses through zero. The associated probability amplitude will thus possess a single radial node, corresponding to a $ 2s $ state. For a superposition with additional Gaussians which alternate in signature as the widths increase, one can produce more nodes in the wave function.
	These techniques of identifying radial excitations by superposing smeared quark sources and sinks in a correlation matrix analysis reflect similar findings from direct calculations of the state wave functions \cite{Roberts:2013ipa,Roberts:2013oea}. 
	This is a qualitative strength of our approach when compared with others in the literature; using combinations of smeared Gaussians allows for powerful insight into the macroscopic structural properties of the state wave functions through invariants like radially symmetric nodes.
	
	\subsection*{Charge Radius of the $ \Omega^- $ Baryon} \label{subsec:OmegaRadius}
	Before presenting our results, we first comment on the use of our lattice volume in studying the $ \Omega $-baryon spectrum. The approximate size of the $ \Omega^- $ ground state can be estimated by studying its electromagnetic charge radius. In Ref.~\cite{Boinepalli:2009sq} this was found, using quenched QCD on the lattice, to be $ \langle r^2_{\Omega^-} \rangle = 0.307\pm 0.015\ \mathrm{fm}^2 $. Taking the square root and doubling to get a diameter, we find
	\begin{equation}
		D_\Omega = 1.108 \pm 0.027\ \mathrm{fm}\,.
	\end{equation}
	An updated estimate of the charge radius in full QCD was given in Ref.~\cite{Alexandrou:2010jv} having a value of $ \langle r^2_{\Omega^-} \rangle = 0.348 \pm 0.052\ \mathrm{fm}^2$ at the physical point. This gives a diameter of
	\begin{equation}
		D_\Omega = 1.18 \pm 0.15\ \mathrm{fm}\,.
	\end{equation}
	Both of these results fit comfortably within our lattice volume ($ L\sim3 $ fm in each spatial dimension). This means there is one less obstacle in comparing our mass spectra directly with experimental results. 
	
	\section{Results} \label{sec:OmegaResults}
	Here we present our results for the $ \Omega $-baryon spectrum obtained using the techniques outlined in Section~\ref{sec:Method}. Leveraging a variational approach incorporating spin- and parity-projection operators, we are able to extract three energy eigenstates for each pair of spin and parity quantum numbers $ J^P = 3/2^\pm,\, 1/2^- $, and two states in the $ 1/2^+ $ channel. Each set of results is presented in Figures \ref{fig:omega32+}, \ref{fig:omega32-}, \ref{fig:omega12-} and \ref{fig:omega12+}.
	
	\subsection{Spin-3/2, Even Parity Spectrum}
\begin{figure}
		\centering
		\includegraphics[width=0.8\linewidth]{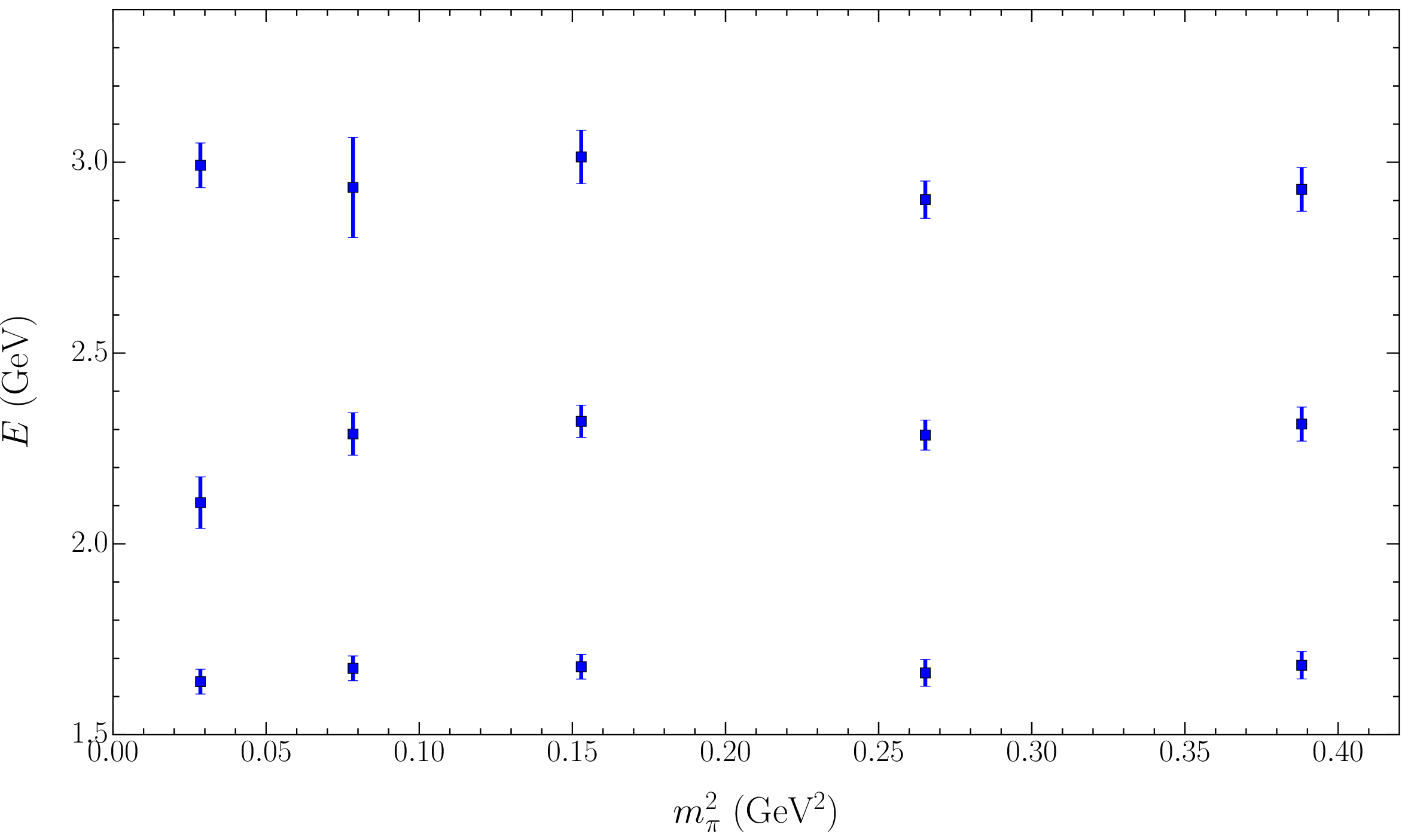}
		\caption{Spectrum of $ \Omega^-(3/2^+) $ baryons across a range of light quark masses quantified by $ m_\pi^2 $.}
		\label{fig:omega32+}
	\end{figure}

	Starting with the $ 3/2^+ $ results in \Fref{fig:omega32+}, in the ground state we see a fairly weak dependence on the light quark masses as we traverse $ m_\pi^2 $. This is also apparent in \Tref{tab:omega_masses} where the column labelled $ m_0 $ shows the ground state masses. We find that each of these is consistent with the experimental value of $ m_\Omega|_\mathrm{expt} = 1.672 $ GeV aside from the result at the lightest pion mass, which sits well below this value. In fact, across several quantum number pairings, there is a tendency for the effective masses in the lightest mass ensemble to lie low in the spectrum. This is caused by the variation in the physical strange quark mass used in the PACS-CS ensembles as the light-quark masses change in the Sommer scheme for setting the lattice spacing, $ a $. While $ \kappa_s $ is held fixed, small changes in $ a $ correspond to small changes in $ m_s $. These changes in $ m_s $ will induce a small slope in our results, and a linear extrapolation is required to accurately obtain results at the physical point.
		
	In extrapolating our masses to the physical point, we use the function,
	\begin{equation}
		m_\Omega (m_\pi^2) = m_\Omega|_\mathrm{phys} + \alpha_2\, \big( m_\pi^2 - m_\pi^2 |_\mathrm{phys}\big)
		\,, \label{eq:mass_extrap}
	\end{equation}
	and fit the parameters $  m_\Omega|_\mathrm{phys} $ and $ \alpha_2  $ to our low-lying lattice results using a $ \chi^2/\mathrm{dof} $ minimsation approach. While this ignores chiral effects at near-physical pion masses, we expect these to be small, and a first order approximation such as that in \Eref{eq:mass_extrap} is sufficient for describing the $ 3/2^+ $ ground state. Initial estimates of the errors in the fit parameters due to extrapolation are obtained by a covariance matrix analysis.
	
	Notably each lattice QCD state in the spectrum shows a fairly constant trend in $ m_\pi^2 $ with a typical drop-off in the mass at the lightest pion mass. This is accounted for in the errors on our extrapolated masses by including a contribution which measures the average deviation of the lightest and second-lightest enesemble results from the line of best fit. This systematic error is averaged and added in quadrature with the extrapolation error, and gives some measure of the scatter of the lightest lattice points about the line of best fit. 
	
	In some instances small errors on lattice results at heavier pion masses lead to errors on the extrapolated masses being smaller than the error on the lattice result closest to the physical point. To avoid underestimating the uncertainty in these cases, we enhance the error on the extrapolated mass to equal the error on the lattice point on the lightest ensemble. 
	
\begin{figure}
		\centering
		\includegraphics[width=0.8\linewidth]{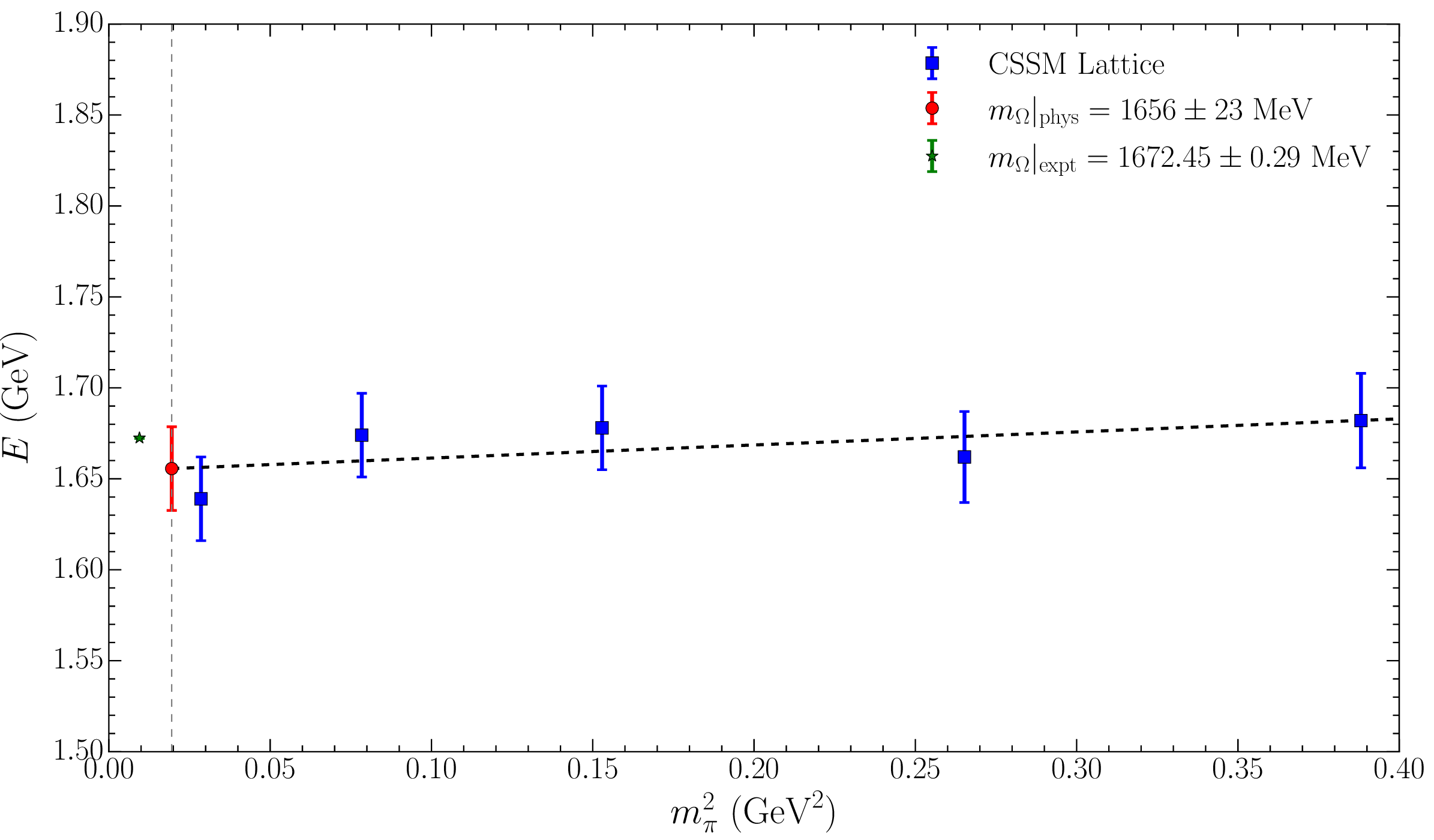}
		\caption{Linear fit of the $ \Omega $ mass function to our lattice QCD results in the $ 3/2^+ $ spectrum. The line of best fit is given by the dashed black line, which fits the lattice results shown as blue squares. The physical pion mass point is indicated by the vertical grey dashed line. Our extrapolation of the $ \Omega $ mass to the physical point is given by the red data point, to be compared with the experimentally observed $ \Omega^-(1672) $ mass shown by the green star (offset for clarity).}
		\label{fig:omega32+_extrap_linear}
	\end{figure}

	In the case of the $ 3/2^+ $ ground state, performing the extrapolation to the physical point gives a mass of
	\begin{equation}
		m_{\Omega}^{3/2^+}|_\mathrm{phys} = 1656 \pm 23\ \mathrm{MeV}\,. \label{eq:omega_mass_extrap}
	\end{equation}
	The corresponding fit is plotted in \Fref{fig:omega32+_extrap_linear}. This is in agreement both with the value of $ m_\Omega = 1650(20) $ in Ref.~\cite{Engel:2013ig}, and with the PDG value of $ m_\Omega|_\mathrm{expt} = 1672.45\pm0.29 $ MeV.
	
	Recent precision results for the spin-3/2 $ \Omega $ baryons were made available in Ref.~\cite{Hudspith:2024kzk} across a range of $ m_\pi^2 $ values. At their pion mass closest to the physical point with $ m_\pi \sim 131 $~MeV, they find $ m_\Omega = 1663.58\pm3.58 $ MeV, which is also in agreement with our determination of the $ \Omega $ mass at the physical point. 
	
	\csvstyle{myTableStyle}{
		tabular=llllllllll,
		head=false,
		table head= \\ \br, 
		late after first line=\\ \mr, 
		table foot=\br, 
	}
		
	\begin{table}
		\caption{
			\label{tab:omega_masses}
			Table of effective mass fit values for the $ \Omega^-(3/2^+) $ spectrum, in units of GeV, across all five pion masses $ m_\pi $. For each of the states extracted having state numbers $ 0, 1, 2 $, we obtain masses $ m $ with uncertainties $ \Delta m $ and chi-squared per degree of freedom values $ \chi^2/\mathrm{dof} $.
		}
		\begin{indented}
			\item[]
			\csvreader[myTableStyle]{./omega32+.in}{}{\csvcoli & \csvcolii & \csvcoliii & \csvcoliv & \csvcolv & \csvcolvi & \csvcolvii & \csvcolviii & \csvcolix & \csvcolx}
		\end{indented}	
		
	\end{table}
	
\begin{figure}
		\centering
		\includegraphics[width=0.8\linewidth]{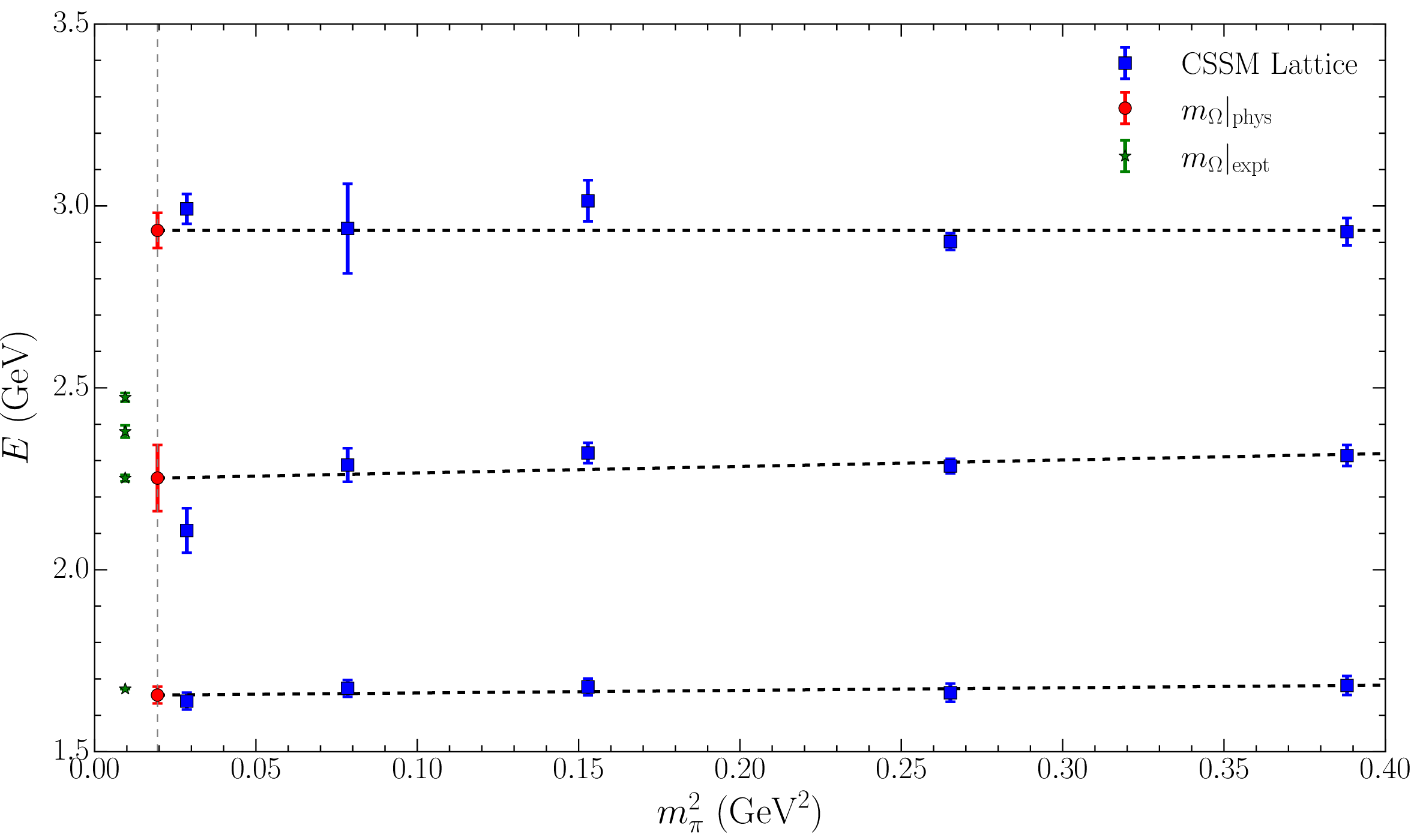}
		\caption{Plot of our results for masses in the $ 3/2^+ $ spectrum, shown as blue squares against $ m_\pi^2 $. The best fit line assuming a linear ansatz is shown by the black dashed lines, and extrapolated to the physical point (grey vertical dashed line). The values of the $ \Omega $ masses at the physical point are shown by the red circles, to be compared with the experimental observations of resonances (green stars offset for visibility), which may have even parity.}
		\label{fig:omega_32+_linear}
	\end{figure}

	We have performed the same fitting procedure for the excited state results with the fits to all three energy levels combined in \Fref{fig:omega_32+_linear}. We've also included the experimental observations of resonances, offset from the physical point for clarity. There is a strong indication that the ground state observed on the lattice is indeed the $ \Omega^-(1672) $, and the first excitation on the lattice coincides nicely with the $ \Omega^-(2250) $. Finally, we note an approximate energy splitting of about $ 600-700 $ MeV between energy levels in the extrapolated masses, across all but the lightest ensemble. This has important implications in terms of an underlying quark model, as has been observed in the $ \Delta $-baryon spectrum \cite{Hockley:2023yzn}.
	
	\subsection{Spin-3/2, Even Parity Radial Excitations}
	We emphasise that our approach to determining the masses in the spectrum is well-suited to identifying radial excitations. As in Ref.~\cite{Hockley:2023yzn}, this is done based on identifying nodes of the wave functions used in exciting these states. This is a key distinguishing feature of our study from others in the literature.
	
\begin{figure}
		\centering
		\includegraphics[width=0.8\linewidth]{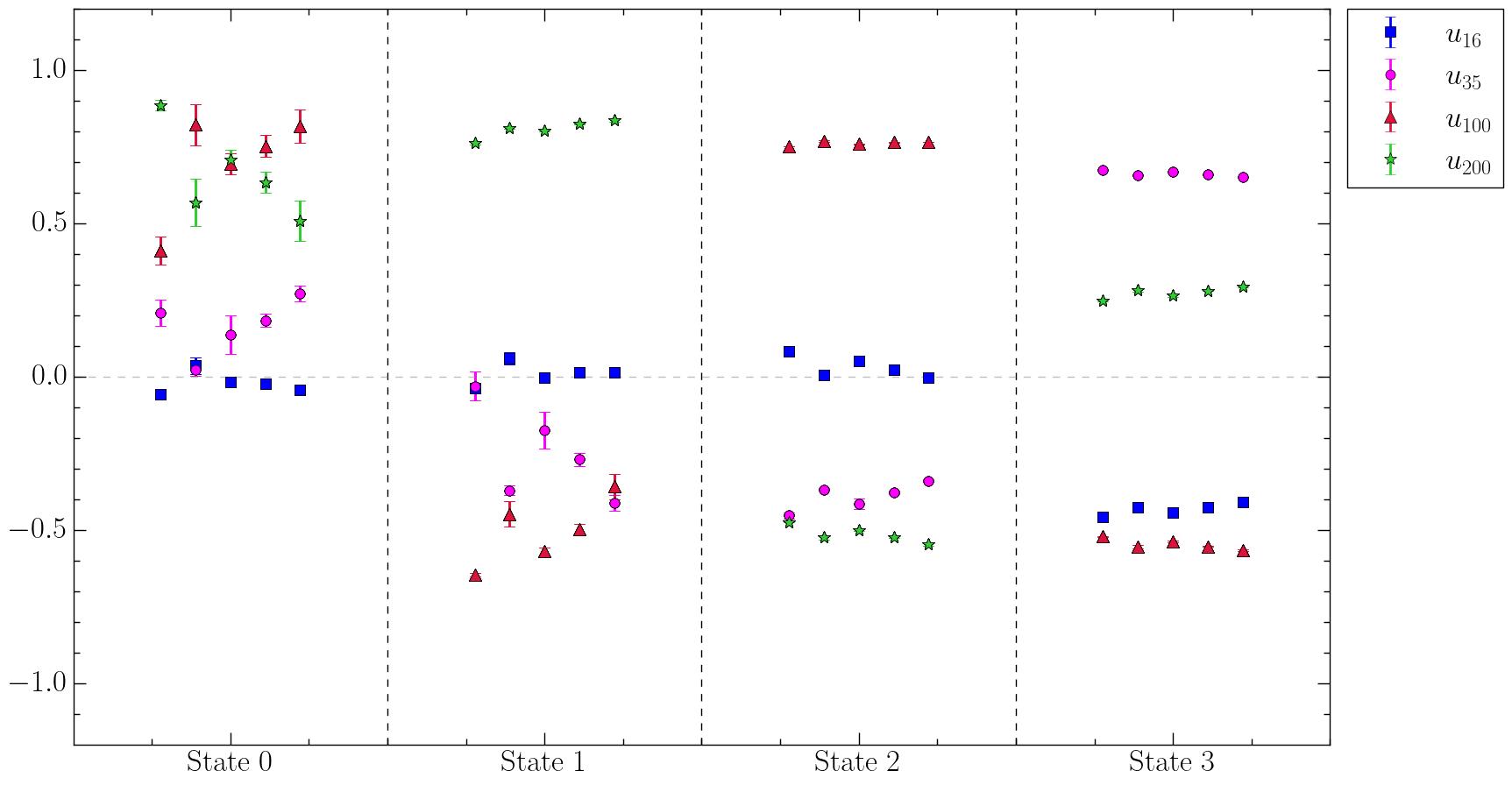}
		\caption{Plot of the $ u^\alpha_i $ components for energy eigenstates $ \alpha = 0, 1, 2, 3 $ and with smearing numbers $ i = 16, 35, 100, 200 $ in the $ 3/2^+ $ spectrum. Quark masses increase from left to right within each column separated by the vertical dashed lines.}
		\label{fig:uvec_omega32+}
	\end{figure}

	We begin by presenting the components of the $ \boldsymbol{u}_\alpha $ vectors for each energy eigenstate $ \alpha $, where $ \boldsymbol{u} $ is a vector of optimised weights for each of the smeared interpolating fields used in our analysis. These eigenvectors are defined by first scaling the weight of each interpolating field such that the diagonal entries of the correlation matrix take the value of 1, at one time slice after the source position. This ensures the elements of the eigenvectors are of $ \mathcal{O}(1) $. These normalised weights are given in \Fref{fig:uvec_omega32+}. We find that, as the number of smearing sweeps increases, the components in state $ \alpha $ cross through zero $ \alpha $ times. In other words, we identify our ground state as a $ 1s $ state, with the corresponding $ 2s $ radial excitation in the region of $ \sim 2.2 $ GeV. The $ 3s $ radial excitation appears at $ \sim 2.9 $ GeV.
	
\begin{figure}
		\centering
		\subfloat[State 0]{ \includegraphics[width=0.31\linewidth]{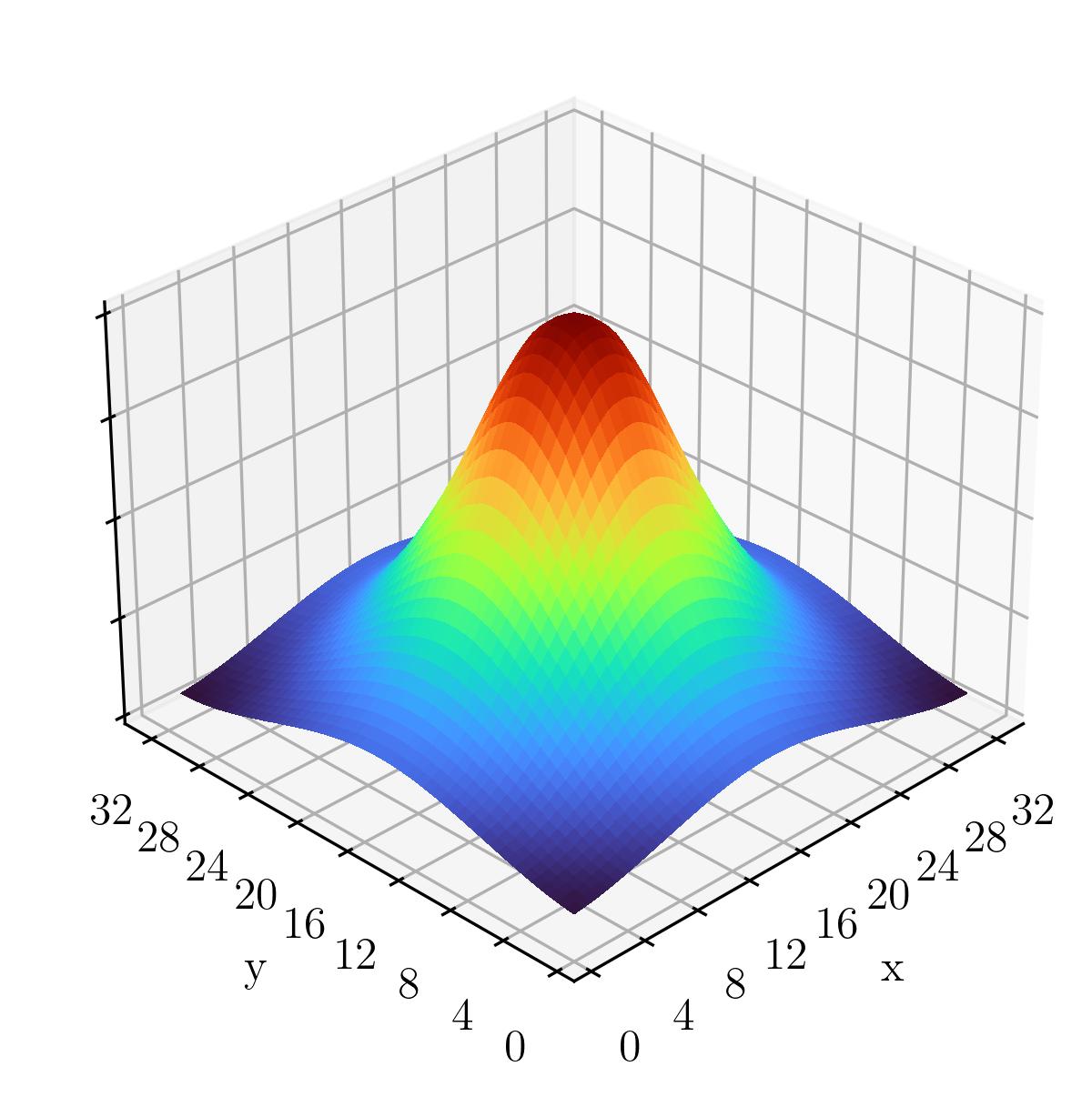}
		}\label{fig:omega32+state0}
		\subfloat[State 1]{ \includegraphics[width=0.31\linewidth]{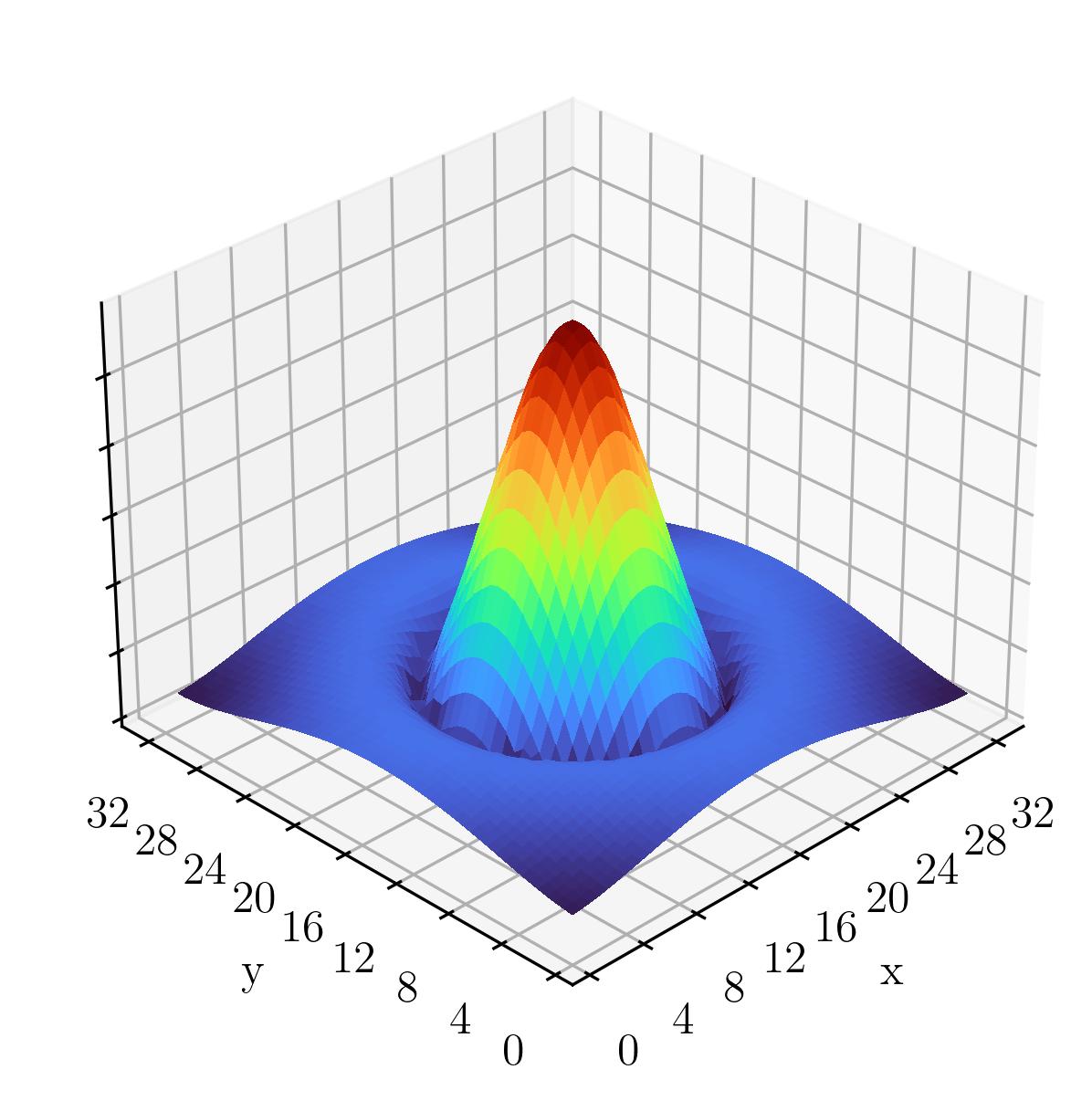}
		}\label{fig:omega32+state1}
		\subfloat[State 2]{ \includegraphics[width=0.31\linewidth]{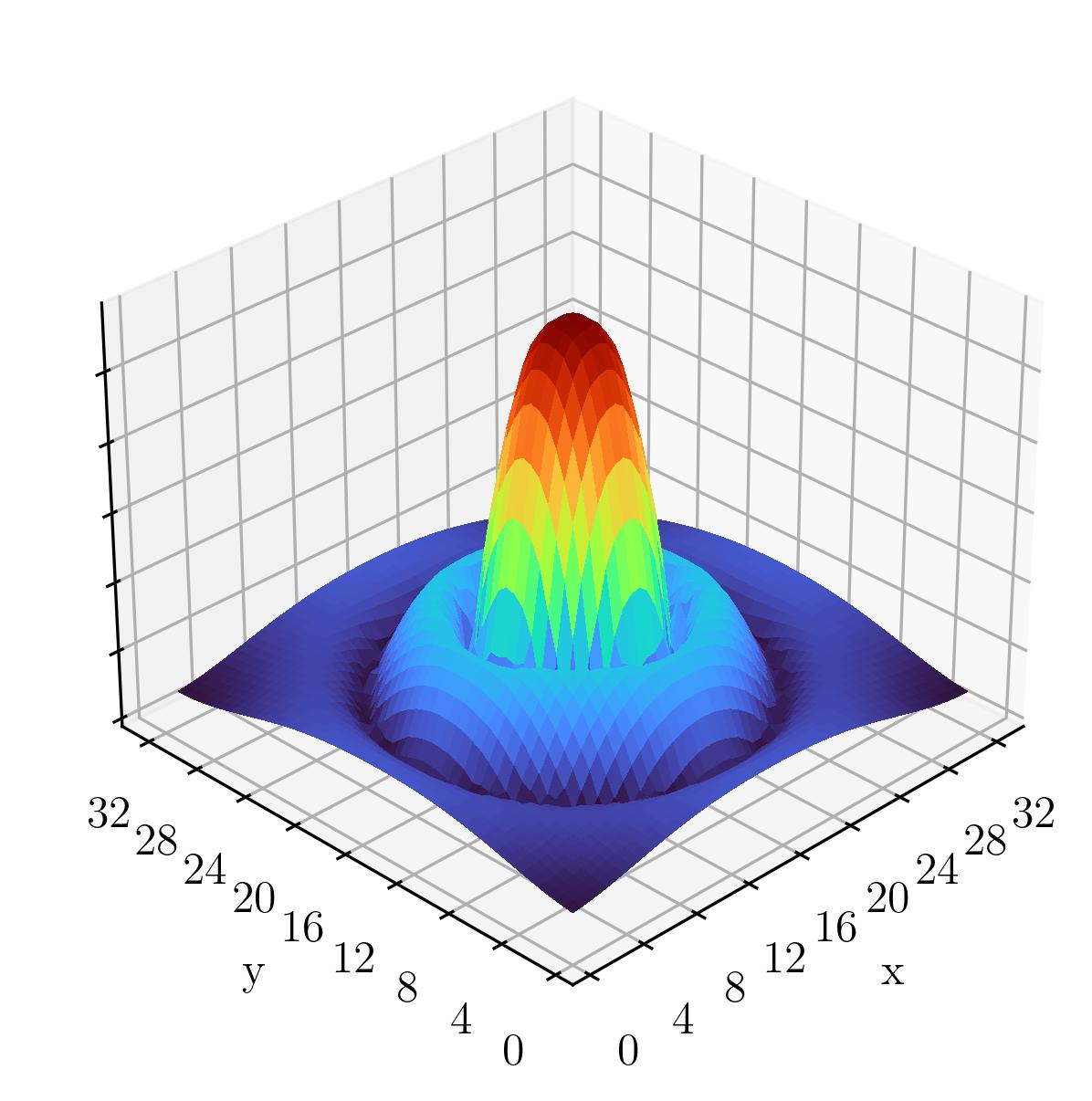}
		}\label{fig:omega32+state2}
		\caption{Probability amplitude distributions for the $ 3/2^+ $ $ \Omega $-baryon spectrum, constructed	from the superposition of smeared interpolating fields, based on the middle quark mass ensemble. These are obtained by taking a superposition of normalised smeared sources
			with weights given by the $ \boldsymbol{u} $ vector components in \Fref{fig:uvec_omega32+}. Nodes in the wave function are indicated by radially symmetric regions of dark blue away from the edges of the volume. (a) No nodes are present in state 0, a $ 1s $ state. (b) A single radially symmetric node is present in state 1, a $ 2s $ state. (c) An inner and outer node are visible in state 2, a $ 3s $ state.}
		\label{fig:nodes_omega32+}
	\end{figure}

	The nodes in the GEVP wave function used for exciting each of the three states reported are depicted in \Fref{fig:nodes_omega32+}. This gives a clear indication of the presence of nodes, reminiscent of both the $ \Delta $ spectrum study in Ref.~\cite{Hockley:2023yzn}, and previous studies of the first positive parity radial excitation of the nucleon \cite{Roberts:2013ipa,Roberts:2013oea}.
	
	\subsection{Odd-Parity Spectrum}
	Just as in the previous section, we present here the results for the spectrum and node identification of the odd-parity states. This is of particular interest since the nature of the $ \Omega^-(2012) $ resonance is contentious.
		
\begin{figure}
		\centering
		\includegraphics[width=0.8\linewidth]{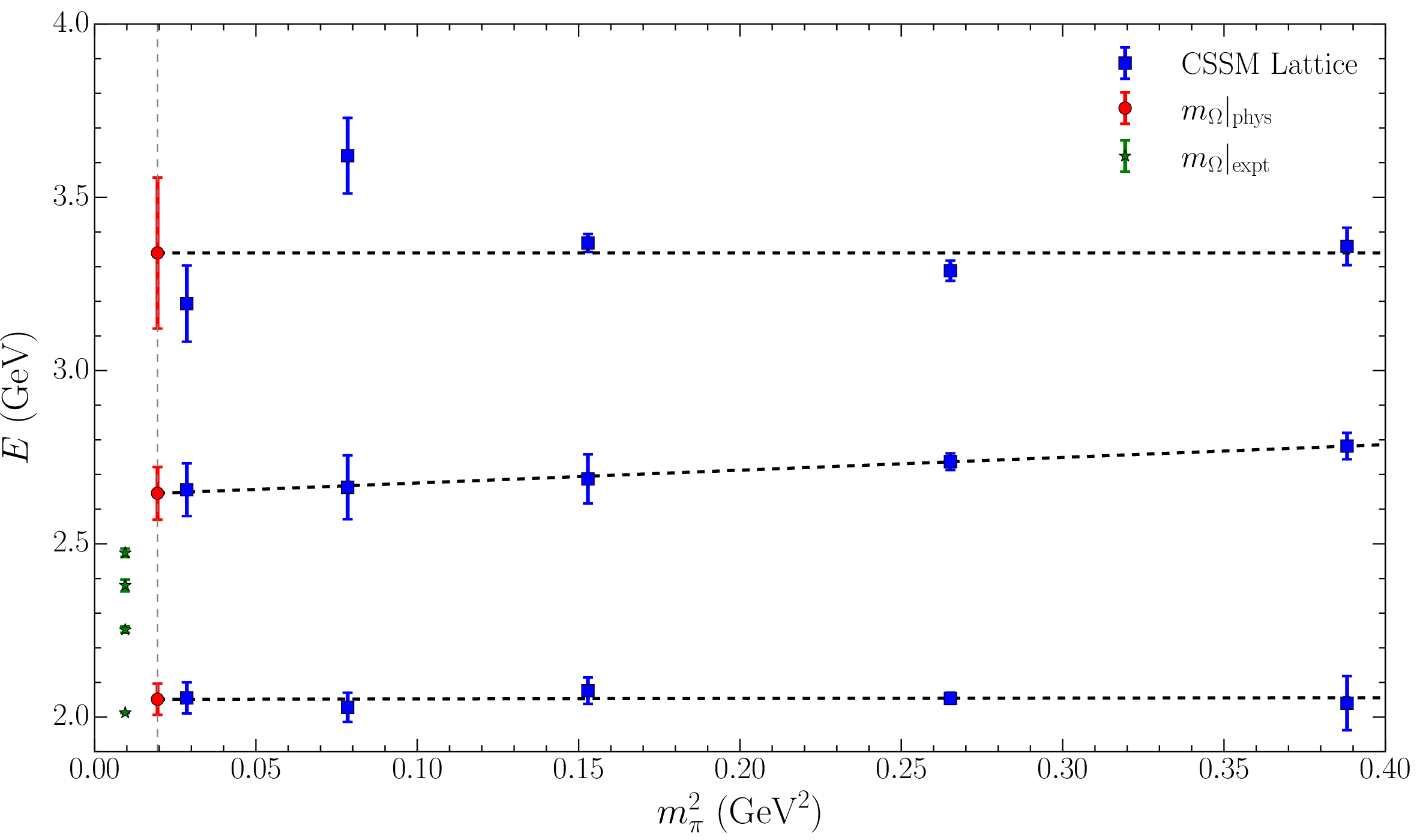}
		\caption{Spectrum of $ \Omega^-(3/2^-) $ baryons across a range of light quark masses quantified by $ m_\pi^2 $ as in \Fref{fig:omega32+}. The result of linear extrapolations to the physical point are shown by red circles.}
		\label{fig:omega32-}
	\end{figure}

\begin{figure}
		\centering
		\includegraphics[width=0.8\linewidth]{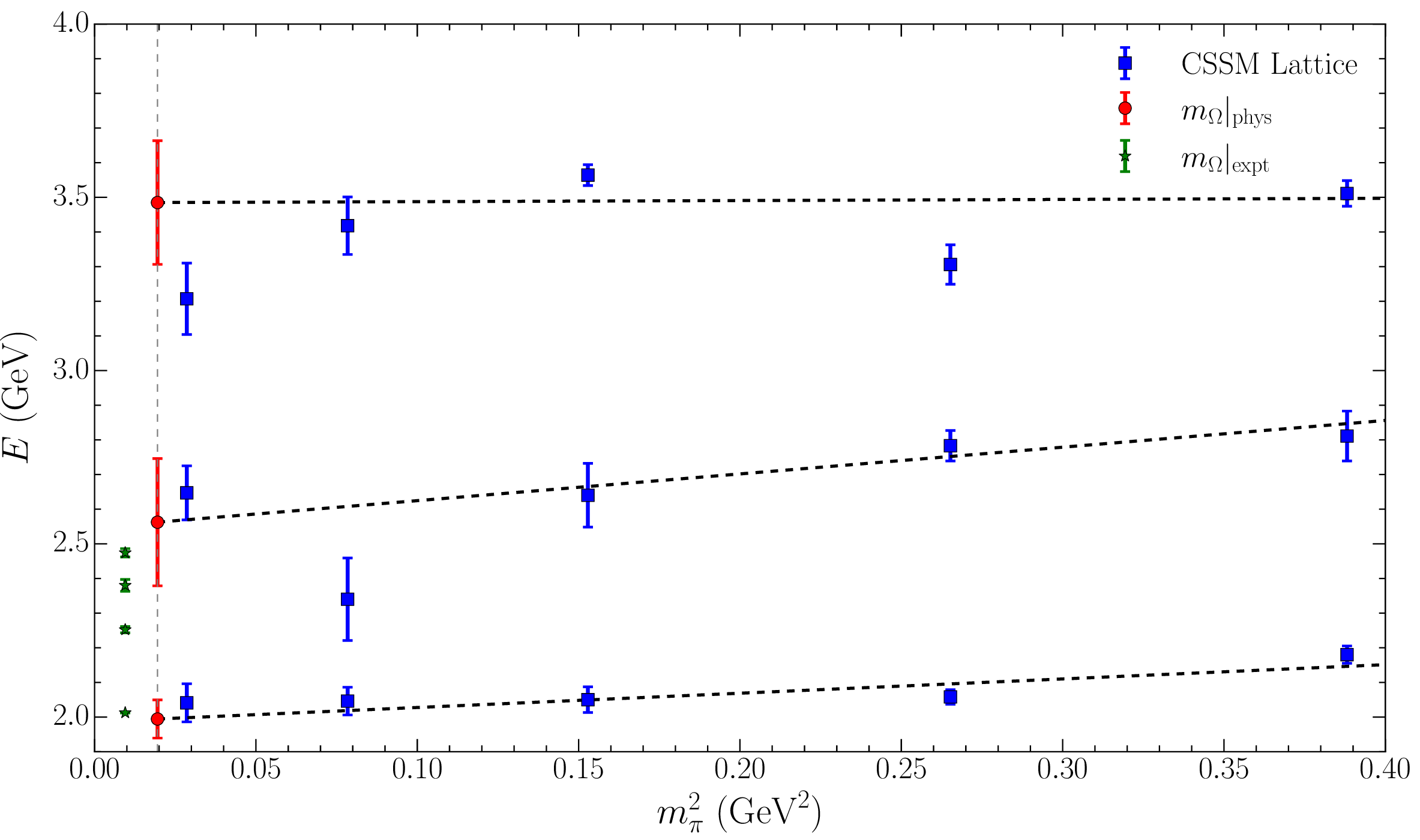}
		\caption{Spectrum of $ \Omega^-(1/2^-) $ baryons across a range of light quark masses quantified by $ m_\pi^2 $ as in \Fref{fig:omega32+}. The results of linear extrapolations to the physical point are shown by red circles.}
		\label{fig:omega12-}
	\end{figure}

	We provide our spectrum results for the spin-3/2 and spin-1/2 odd-parity $ \Omega $-baryons in Figures~\ref{fig:omega32-} and \ref{fig:omega12-} respectively, with the extrapolations to the physical point overlayed. The immediate takeaway is that \textit{both} ground states are consistent with the experimentally measured $ \Omega^-(2012) $ mass. This is remarkably similar to the result found in Ref.~\cite{Engel:2013ig}.
	
	Performing the same mass extrapolation to the physical point, we find, in the spin-3/2 case for the ground state
	\begin{equation}
		m^{3/2^-}_{\Omega}|_\mathrm{phys} = 2051 \pm 45\ \mathrm{MeV}\,,
	\end{equation} 
	and for the spin-1/2 odd-parity ground state
	\begin{equation}
		m^{1/2^-}_{\Omega}|_\mathrm{phys} = 1994 \pm 55\ \mathrm{MeV}\,.
	\end{equation} 
	
	Both of these results are in agreement with the experimental observation of the $ \Omega^-(2012) $ mass, within one standard deviation. Furthermore, they are in excellent agreement with the findings of Ref.~\cite{Engel:2013ig}, with the $ 3/2^- $ result having greater energy than the $ 1/2^- $.
	
	Interestingly, where the ground states of both the $ 3/2^+ $ and $ 3/2^- $ were almost constant in $ m_\pi^2 $, the fit shown for the $ 1/2^- $ ground state in \Fref{fig:omega12-} shows a markedly positive slope. This is caused by the result for the heaviest quark mass ensemble, which sits well above the other results for this energy level. Thus in extrapolating to the physical point, we also consider a fit to the first 4 lightest ensemble results. This yields a revised fit value for the mass at the physical point of
	\begin{equation}
		m^{1/2^-}_{\Omega}|_\mathrm{phys} = 2041 \pm 55\ \mathrm{MeV}\,,
	\end{equation} 
	which is still in agreement with the experimental value up to one standard deviation. 
	
	With both the $ 1/2^- $ and $ 3/2^- $ ground state results being consistent with the experimentally measured $ \Omega^-(2012) $ mass, our results agree with an odd-parity state. This suggests the presence of two overlapping odd-parity resonances contained within the PDG's report of the $ \Omega^-(2012) $ resonance.
	
	In Ref.~\cite{Hudspith:2024kzk} the ensemble with pion mass closest to the physical point has $ m_\pi = 131 $ MeV, and yields a mass of $ m^{3/2^-}_\Omega = 2012 \pm 12 \ \mathrm{MeV} $. This is in agreement with our result for the $ 3/2^- $ ground state. 
	
	Moving on to the excited states, we find strong evidence that there are no other spin-3/2 odd parity resonances in the range $ \sim 2.2-2.5 $ GeV, with our first and second excited states sitting well above the highest energy resonances reported in the PDG. 
	
	The first $ 1/2^- $ excitation may be associated with one of the three PDG resonances in the range $ \sim 2.2-2.5 $ GeV. The second excitation sits well above the observed resonances.
	
	In summary, we find general agreement between our results and those presented in Refs.~\cite{Engel:2013ig,Hudspith:2024kzk}, for the ground states. We find a clear indication of two odd-parity states which correspond to the experimentally measured $ \Omega^-(2012) $ resonance. Our lattice calculations predict an overlap of $ 1/2^- $ and $ 3/2^- $ states in the single resonance reported in the PDG. {Similar findings from a QCD sum rules approach have been recently reported in Ref.~\cite{Su:2024lzy}. Finally, we find strong evidence that the resonances thus far listed in the PDG do not appear to correspond with the first or second excitations in the spin-3/2 odd-parity spectrum. Interestingly, in the $ 3/2^- $ spectrum we find an approximately equal splitting between energy levels at the physical point and across several ensembles, with $ \sim 600-700 $ MeV splitting, just as we found in the $ 3/2^+ $ spectrum.

	\subsection{Odd-Parity Radial Excitations}
	As usual we can identify the radial excitations based on nodes of the wave functions used to excite the states, finding that state $ \alpha $ has $ \alpha $ nodes for $ \alpha = 0, 1, 2 $. The node structure for the odd-parity states is shown in Figures~\ref{fig:nodes_omega32-} and \ref{fig:nodes_omega12-}. For the odd-parity case, we recall the discussion of Ref.~\cite{Hockley:2023yzn}; the odd-parity states are accessed via the opposite parity lower Dirac components of the lattice correlation functions. As such, the spatial wave functions which multiply the Dirac components are even parity and appear just as for the even-parity states. 
	A process involving creating the odd-parity state and annihilating it with an even-parity Dirac operator is required for realising the parity structure in the spatial wave function that one would associate with say, an odd-$ l $ resonance ($ p $-wave or $ f $-wave for example) \cite{Kamleh:2017lye}.
	
	\begin{figure}
		\centering
		\subfloat[State 0]{ \includegraphics[width=0.31\linewidth]{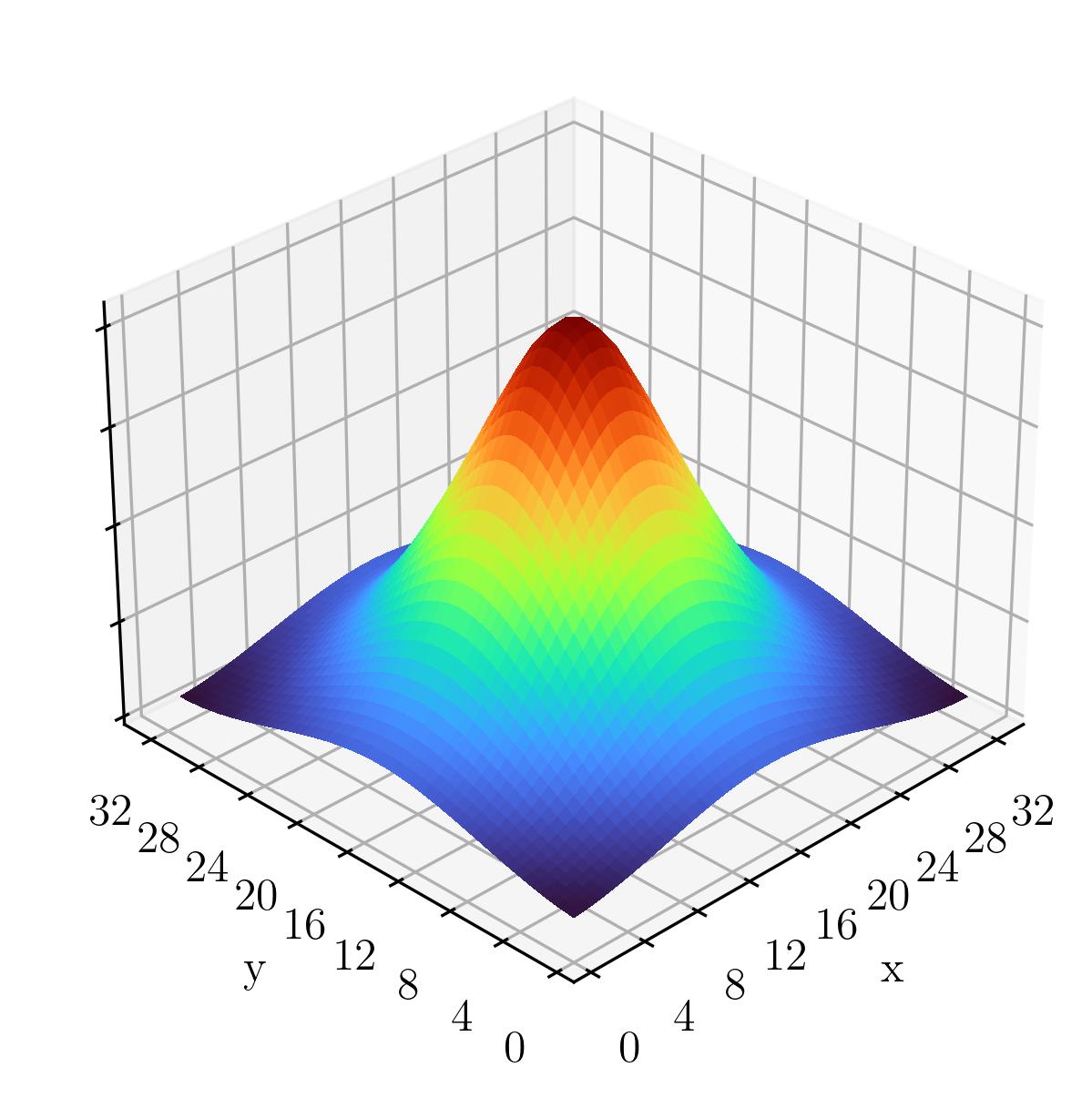}
		}\label{fig:omega32-state0}
		\subfloat[State 1]{ \includegraphics[width=0.31\linewidth]{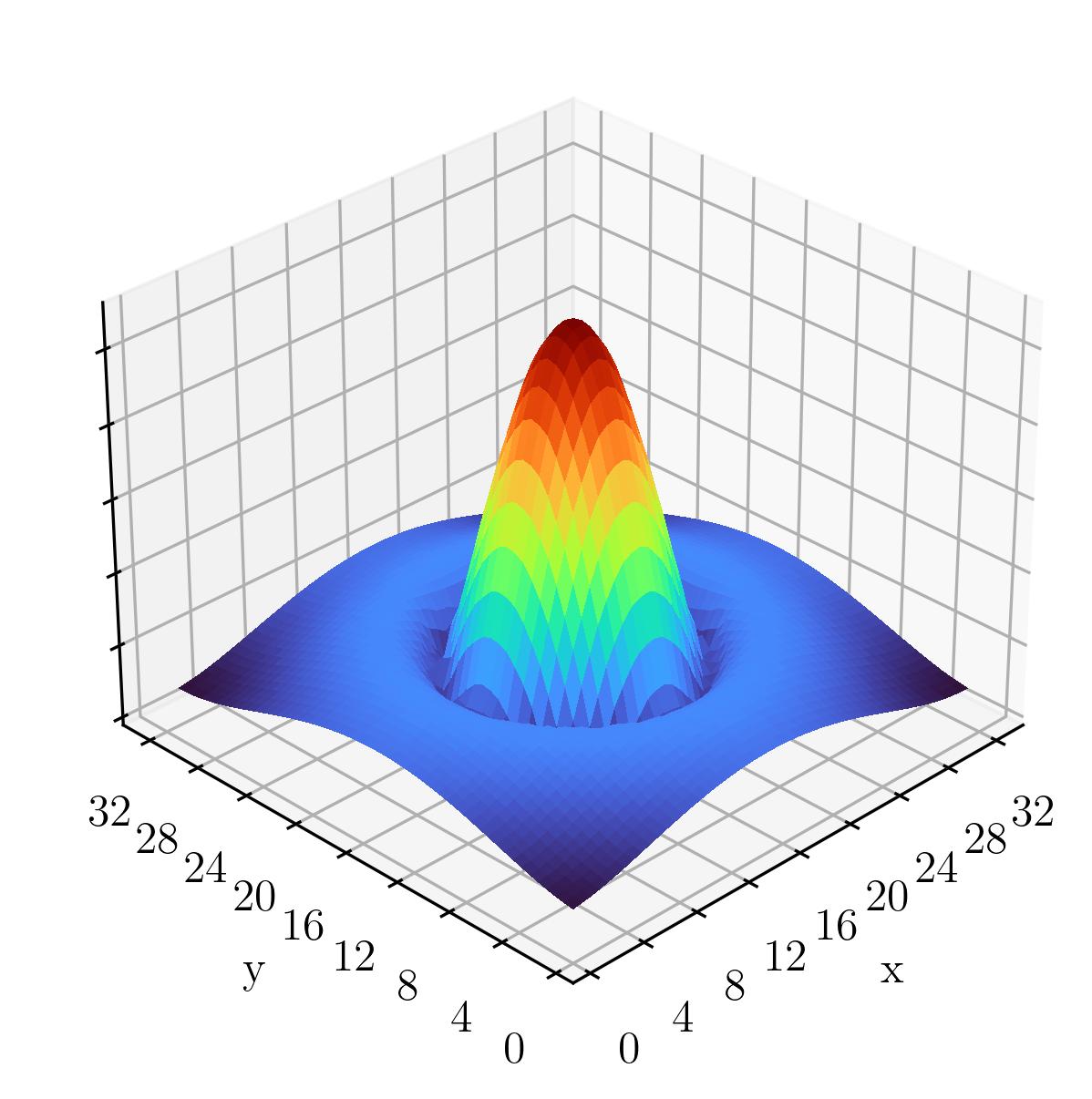}
		}\label{fig:omega32-state1}
		\subfloat[State 2]{ \includegraphics[width=0.31\linewidth]{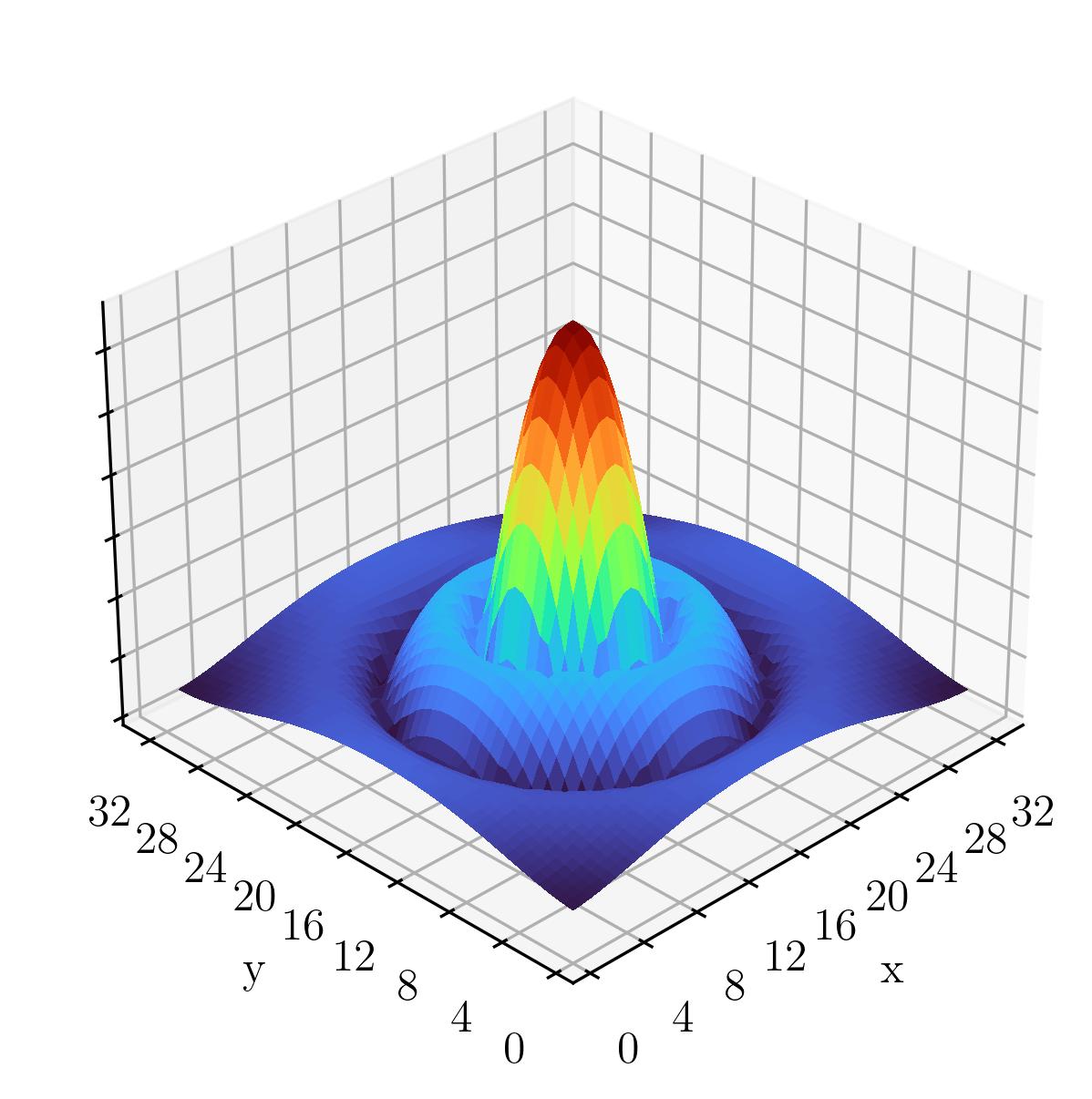}
		}\label{fig:omega32-state2}
		\caption{Same as in \Fref{fig:nodes_omega32+} but for the case of $ J^P = 3/2^- $. (a) No nodes are present in state 0. (b) A single radially symmetric node is present in state 1. (c) An inner and outer node are visible in state 2. Note, these even parity spatial wave functions multiply the lower components of the Dirac operators to create odd-parity interpolators.}
		\label{fig:nodes_omega32-}
	\end{figure}
	
	\begin{figure}
		\centering
		\subfloat[State 0]{ \includegraphics[width=0.31\linewidth]{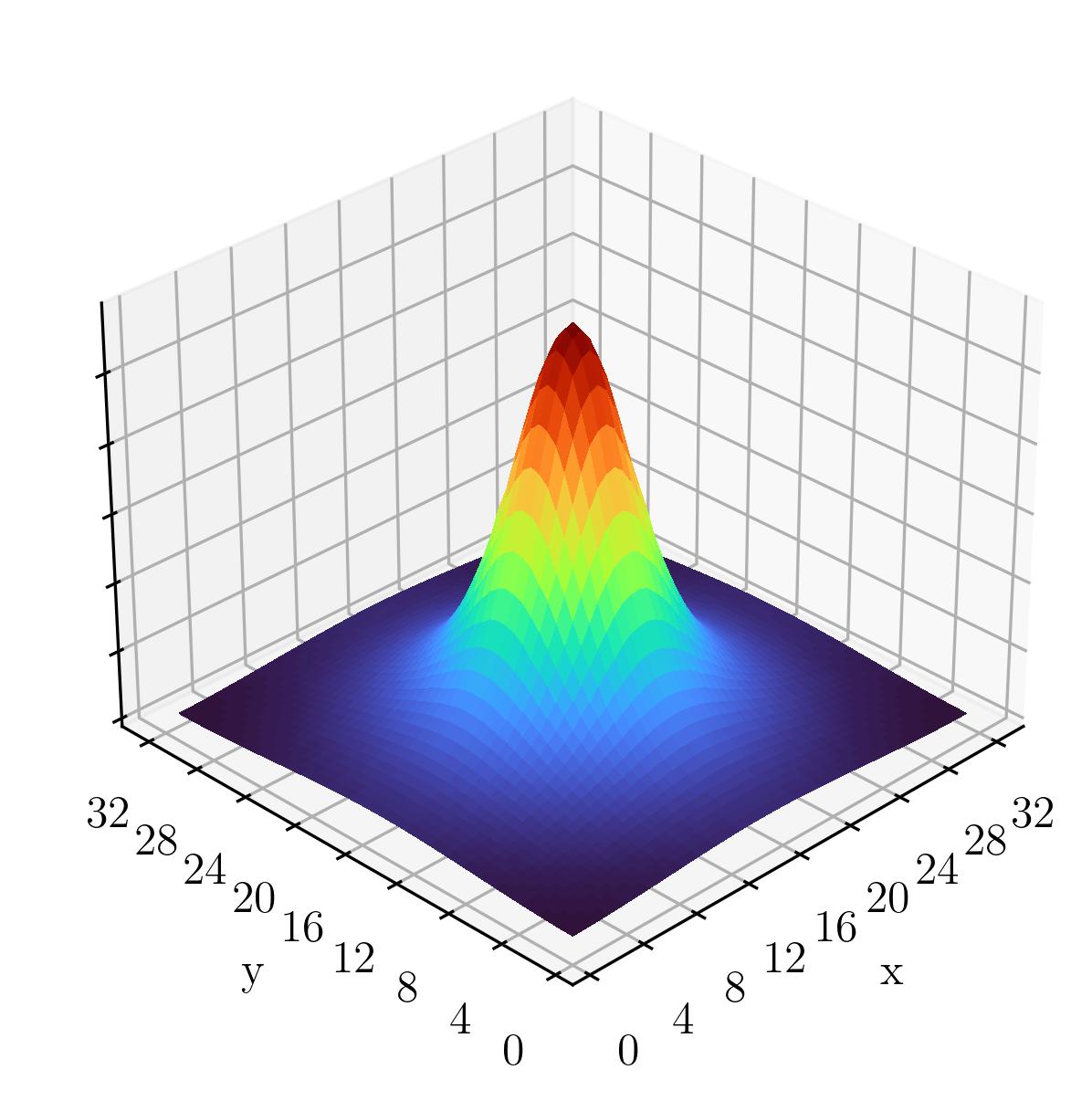}
		}\label{fig:omega12-state0}
		\subfloat[State 1]{ \includegraphics[width=0.31\linewidth]{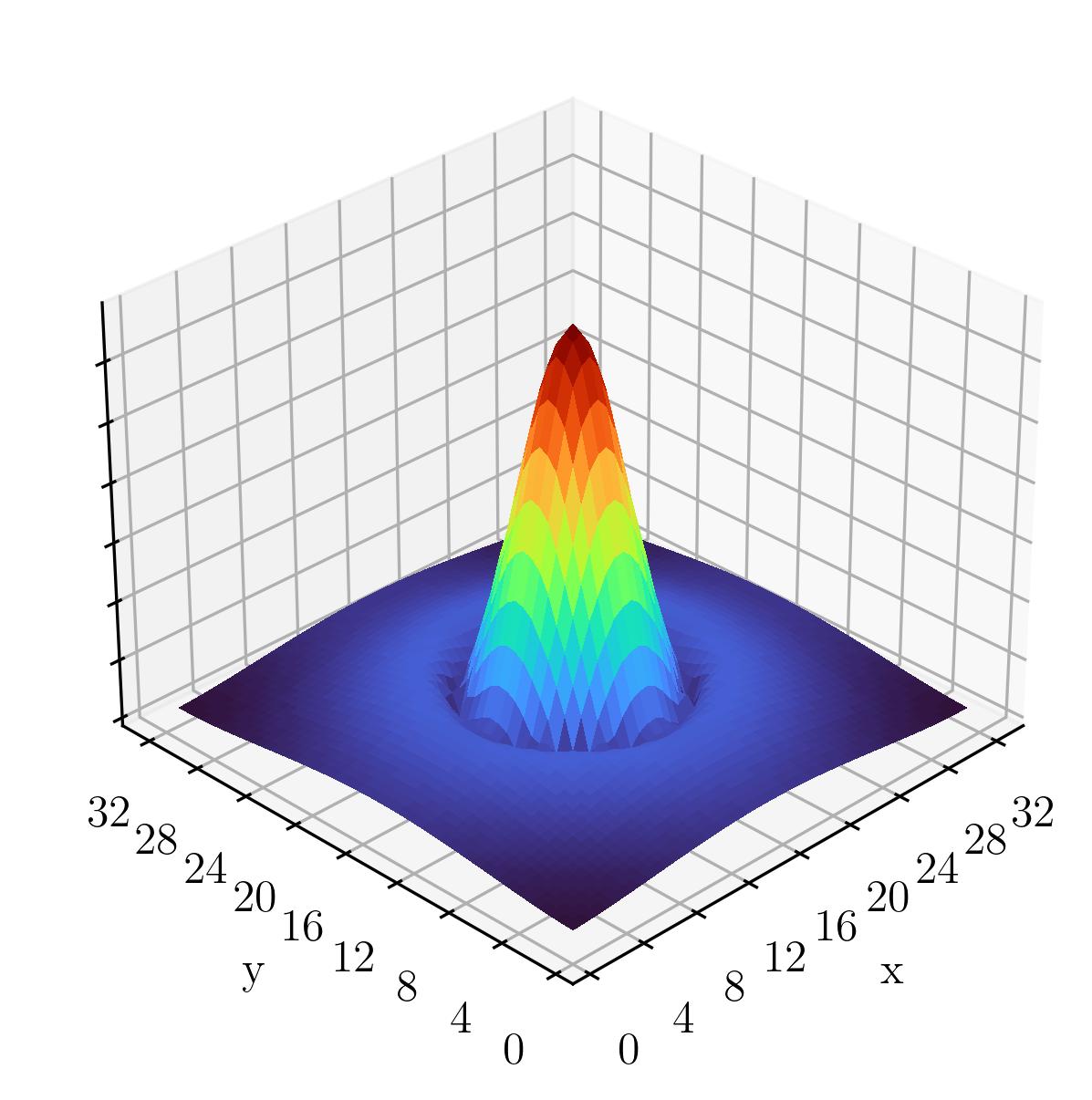}
		}\label{fig:omega12-state1}
		\subfloat[State 2]{ \includegraphics[width=0.31\linewidth]{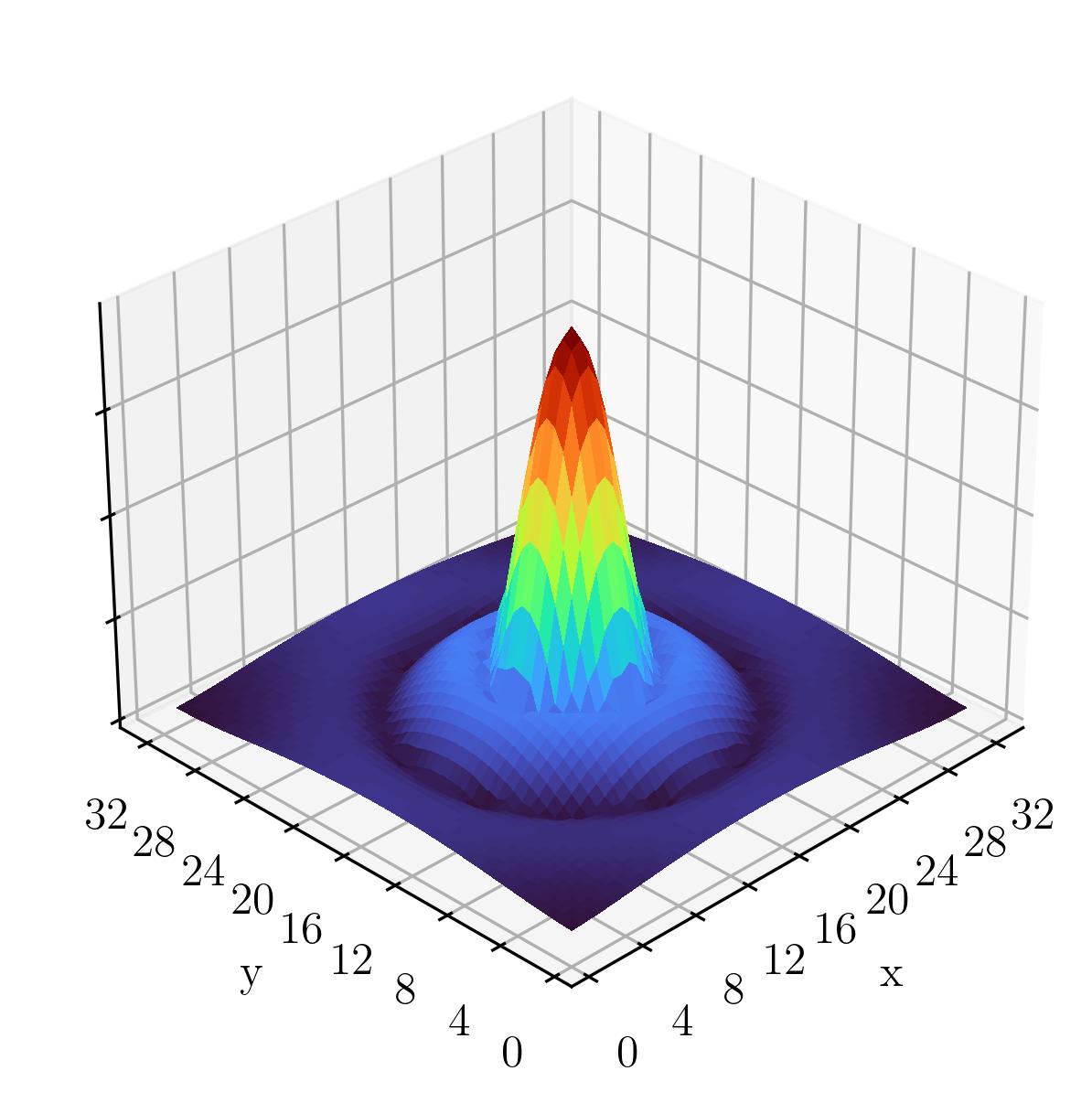}
		}\label{fig:omega12-state2}
		\caption{Same as in \Fref{fig:nodes_omega32+} but for the case of $ J^P = 1/2^- $. (a) No nodes are present in state 0. (b) A single radially symmetric node is present in state 1. (c) An inner and outer node are visible in state 2. Note, these even parity spatial wave functions multiply the lower components of the Dirac operators to create odd-parity interpolators.}
		\label{fig:nodes_omega12-}
	\end{figure}
	
	\subsection{Spin-1/2, Even Parity Spectrum}
	Results for the energies of the ground and first excited states as determined using lattice QCD are given in \Fref{fig:omega12+}. Performing the same extrapolations as before, we find a linearly extrapolated lowest state mass of
	\begin{equation}
		m_{\Omega}^{1/2^+}|_\mathrm{phys} = 2342 \pm 140\ \mathrm{MeV}\,.
	\end{equation}
	
	Owing to the large uncertainty on this result, this has overlap with the three highest energy resonances in the PDG, the $ \Omega^-(2250),\, \Omega^-(2380) $ and $ \Omega^-(2470) $ resonances. The first excitation result is well above the PDG resonances at $ \sim 3 $ GeV.
	
	\begin{figure}
		\centering
		\includegraphics[width=0.8\linewidth]{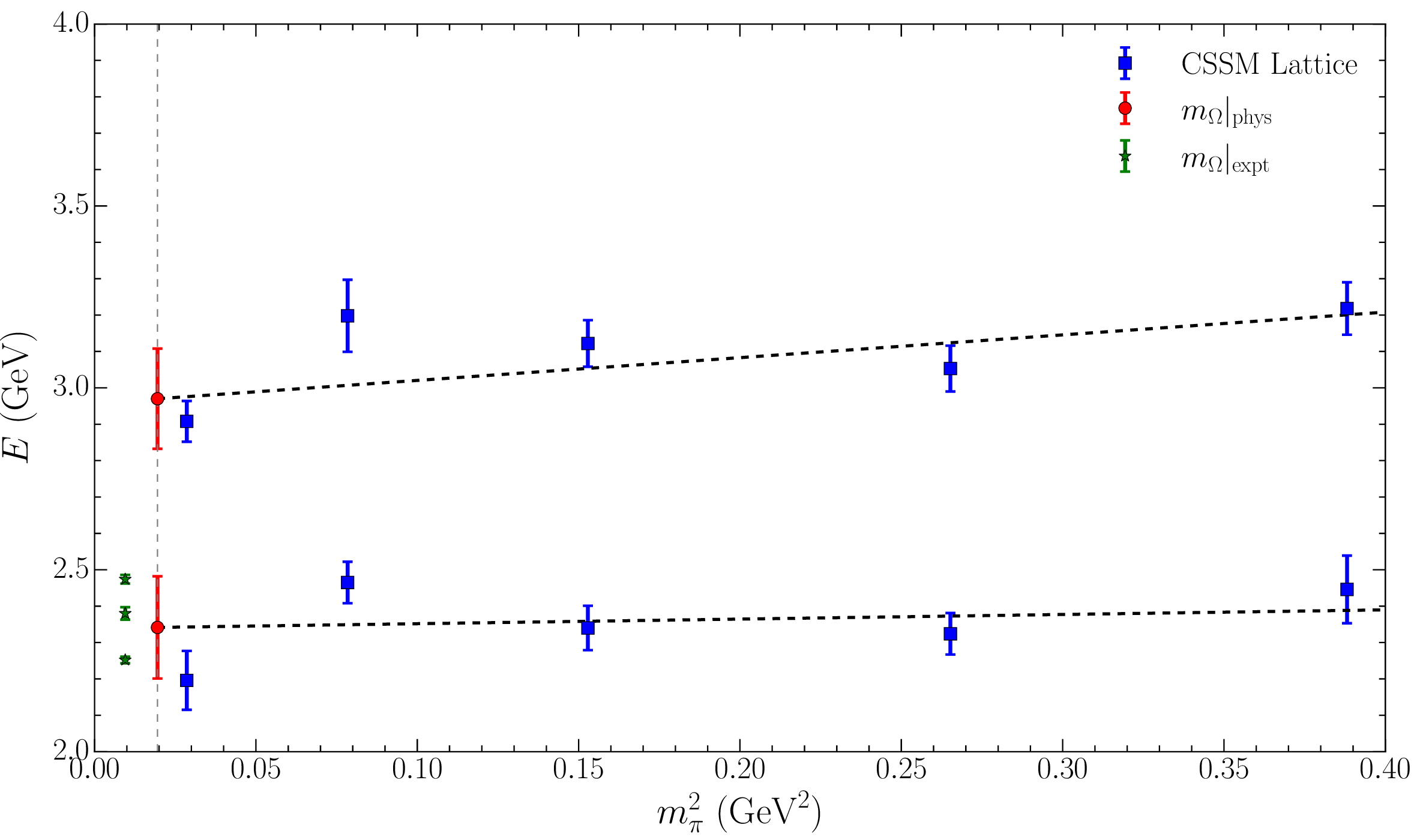}
		\caption{Spectrum of $ \Omega^-(1/2^+) $ baryons across a range of light quark masses quantified by $ m_\pi^2 $ as in \Fref{fig:omega32+}. The results of linear extrapolations to the physical point are shown by red circles.}
		\label{fig:omega12+}
	\end{figure}
	
	\subsection{Spin-1/2, Even Parity Radial Excitations}
	As in the previous sections, we identify the radial excitations of the states in the $ 1/2^+ $ spectrum by counting the nodes of the wave function. The ground state is identified as a $ 1s $ state with the first excitation being the corresponding $ 2s $ radial excitation. The plots of the wave functions used to excite the states in the lattice GEVP are given in \Fref{fig:nodes_omega12+}.
	
	\begin{figure}[h]
		\centering
		\hfill
		\subfloat[State 0]{ \includegraphics[width=0.4\linewidth]{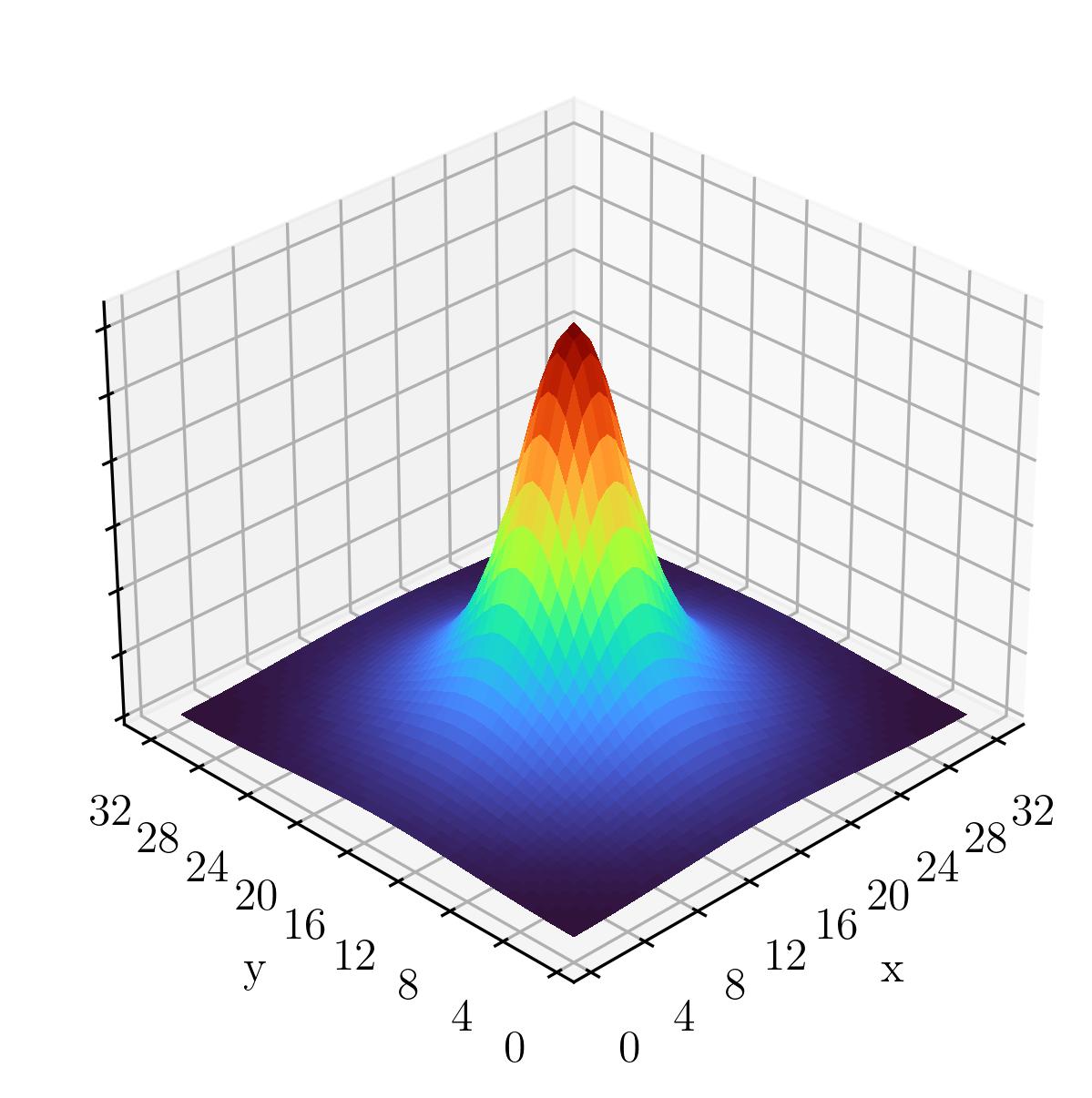}
		}\label{fig:omega12+state0}
		\hfill
		\subfloat[State 1]{ \includegraphics[width=0.4\linewidth]{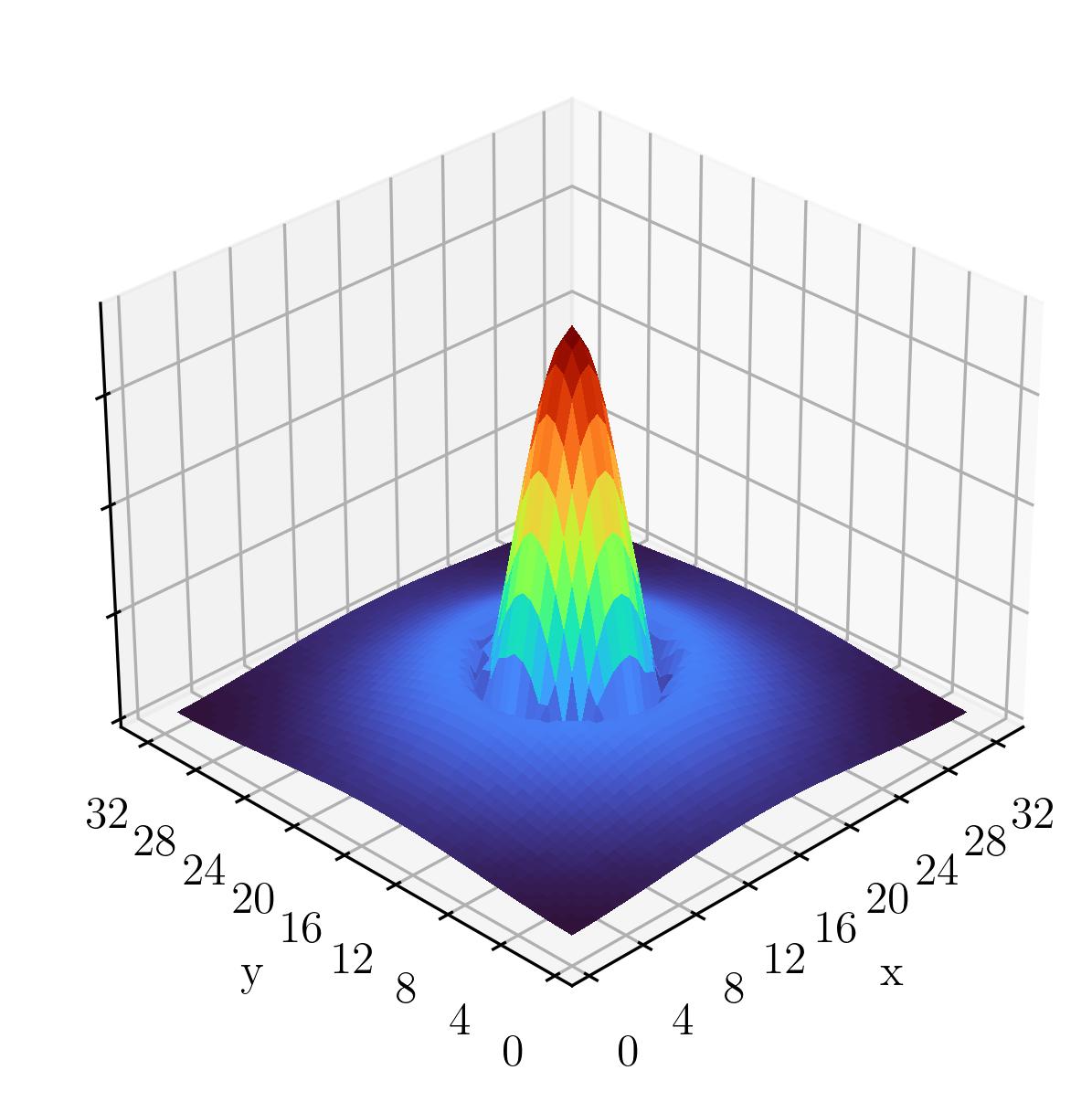}
		}
		\hfill
		\label{fig:omega12+state1}
		\caption{Same as in \Fref{fig:nodes_omega32+} but for the case of $ J^P = 1/2^+ $. (a) No nodes are present in state 0, a $ 1s $ state. (b) A single radially symmetric node is present in state 1, so this is identified as a $ 2s $ state.}
		\label{fig:nodes_omega12+}
	\end{figure}
	
	\section{Remarks}\label{sec:Remarks}
	Thus far, we have attributed the $ \Omega^-(1672) $ to our $ 3/2^+ $ $ 1s $ state on the lattice, and the $ \Omega^-(2012) $ resonance to both of our lowest-lying $ 3/2^- $ and $ 1/2^- $ masses, suggesting the current PDG resonance is actually two. Now we look to see if we can suggest any quantum numbers for the remaining resonances of \Tref{tab:OmegaPDG}. If these are dominated by a three-quark core, we would expect these to be associated with one of the finite-volume eigenstates accessed in the lattice GEVP. The current status of the resonances identified by the PDG is shown in \Tref{tab:omega_extraps_linear} alongside our findings from the lattice calculations conducted. The resonances are placed in the same rows as the lattice results with masses closest to those in the PDG. For the first two resonances which we have already discussed in detail, this association is clear in the case of the $ \Omega^-(1672) $, and the reported $ \Omega^-(2012) $ appears to be associated with both our $ 1/2^- $ and $ 3/2^- $ lattice results.
	
	\begin{table}[]
		\caption{
			\label{tab:omega_extraps_linear}
			Table comparing the PDG masses and quantum numbers $ J^P $ in the $ \Omega $-baryon spectrum with those obtained in this work. The reported lattice masses are the extrapolated $ \Omega $ masses obtained by taking a linear fit in $ m_\pi^2 $ to our lattice QCD results. The Nodes column indicates the number of radial nodes identified in the state wave function.
		}
		\begin{indented}
			\item[]
			\begin{tabular}{llllll}
				\br
				\centre{3}{PDG} & \centre{3}{Lattice} \\ \ms
				\crule{3} & \crule{3} \\ \ms
				Resonance & $ J^P $ & $ m $ (MeV) & $ J^P $ & $ m_\Omega|_\mathrm{phys} $ (MeV) & Nodes \\
				\mr
				$ \Omega^-(1672) $ & $ 3/2^+ $ & $ 1672.5 \pm \hphantom{0}0.3 $ & $ 3/2^+ $ & $ 1656\pm\hphantom{0}23  $ & $ 0 $  	\\
				$ \Omega^-(2012) $ & $ ?^- $   & $ 2012.4 \pm \hphantom{0}0.9 $	& $ 1/2^- $ & $ 2041\pm\hphantom{0}55  $ & $ 0    $	\\
				& 		   &  					& $ 3/2^- $ & $ 2051\pm\hphantom{0}45  $ & $ 0     $	\\
				$ \Omega^-(2250) $ & $ ?^? $   & $ 2252\hphantom{.0} \pm \hphantom{0}9  $ 	& $ 3/2^+ $ & $ 2252\pm\hphantom{0}91  $ & $ 1 $  	\\
				$ \Omega^-(2380) $ & $ ?^? $   & $ 2380\hphantom{.0} \pm 17 $ 	& $ 1/2^+ $ & $ 2341\pm140 $ & $ 0 $  	\\
				$ \Omega^-(2470) $ & $ ?^? $   & $ 2474\hphantom{.0} \pm 12 $ 	& $ 1/2^- $ & $ 2562\pm184 $ & $ 1     $ 	\\
				& 		   &	  				& $ 3/2^- $ & $ 2646\pm\hphantom{0}76  $ & $ 1	   $	\\
				& 		   &    				& $ 3/2^+ $ & $ 2933\pm\hphantom{0}48  $ & $ 2 $  	\\
				& 		   &	  				& $ 1/2^+ $ & $ 2970\pm138 $ & $ 1 $  	\\
				& 		   &	  				& $ 3/2^- $ & $ 3339\pm218 $ & $ 2	 $	  	\\
				& 		   &	  				& $ 1/2^- $ & $ 3485\pm178 $ & $ 2	 $  	\\
				\br
			\end{tabular}
		\end{indented}
	\end{table}
	
	For the $ \Omega^-(2250) $, it is reasonable to associate this with either the $ 2s $ state in the $ 3/2^+ $ spectrum, or the $ 1s $ state in the $ 1/2^+ $ spectrum, as both of the extrapolated masses on the lattice agree with the PDG mass within errors. While the central value of the $ 3/2^+ $ extrapolated mass is in excellent agreement with the PDG, the $ 1/2^+ $ ground state uncertainty is broad enough that we cannot rule out an association with this lattice result. We align the resonances as shown in \Tref{tab:omega_extraps_linear} based on the central values alone, noting that the broad uncertainties would need to be reduced to firmly distinguish between these two cases. For now, the most that can be unambiguously concluded is that there are no nearby odd-parity finite-volume eigenstates close to the mass of the $ \Omega^-(2250) $ resonance. This suggests that this resonance has positive parity. Still, the agreement between the energy of the $J^P = 3/2^+$ radial excitation seen in our lattice calculation with the energy of the $\Omega^-(2250)$ and the tension with the position of the next resonance indicates that the $J^P$ assignment is most likely $3/2^+$.
	
	The $ \Omega^-(2380) $ case has the opposite issue in that we are able to determine its spin but not its parity. Its mass lies in agreement with the second lattice result in the $ 1/2^- $ spectrum, and the $ 1/2^+ $ $ 1s $ result. This indicates that this resonance coincides with a spin-1/2 state. A more extensive investigation of interpolating fields coupling to the spin-1/2 states may allow the ambiguity in the parity to be resolved in a future lattice calculation.
	
	The nature of the $ \Omega^-(2470) $ resonance is also not completely resolved, since there are again multiple lattice results with energies in the vicinity of this resonance (the $ 1/2^+ $ $ 1s $ state and the $ 1/2^- $ first excitation). While the first excitations in both odd-parity channels have similar energies around 2.5 and 2.6~GeV, the narrow errors on the $ 3/2^- $ result indicate there is likely no association with the $ \Omega^-(2470) $. The extrapolated masses for each of these lattice results are in agreement with the corresponding states found in Ref.~\cite{Engel:2013ig}, and indicate a similar energy level ordering for the odd-parity states as observed for the ground states. As the closest finite-volume energy eigenstates are spin-1/2, we may tentatively suggest a spin-1/2 quantum number for the resonance at 2470 MeV, though the parity remains unknown.
	
	In summary, we have identified finite-volume eigenstates with the following correspondences to experiment:
	\begin{itemize}
		\item $ \Omega^-(1672) $ - \textbf{a $ \mathbf{3/2^+} $ ground state.}
		\item $ \Omega^-(2012) $ - \textbf{an overlap of $ \mathbf{3/2^-} $ and $ \mathbf{1/2^-} $ ground states.}
		\item $ \Omega^-(2250) $ - \textbf{an even-parity state.} Likely a $ 3/2^+ $ radial excitation.
		\item $ \Omega^-(2380) $ - \textbf{likely a spin-1/2 state.} A $ 1/2^+ $ ground state or $ 1/2^- $ first excited state. 
		\item $ \Omega^-(2470) $ - \textbf{a spin-1/2 state.} A $ 1/2^- $ first excited state or $ 1/2^+ $ ground state. 
	\end{itemize}
	
	These suggested associations are visualised in \Fref{fig:pdg_comparison} where we show our lattice results which lie nearby to PDG resonances. By considering each resonance along the central column, one can quickly deduce the parity by scanning either left or right of the resonance, to see if there are matching lattice results in the odd- or even-parity sectors respectively.
	
	\begin{figure}
		\centering
		\includegraphics[width=0.8\linewidth]{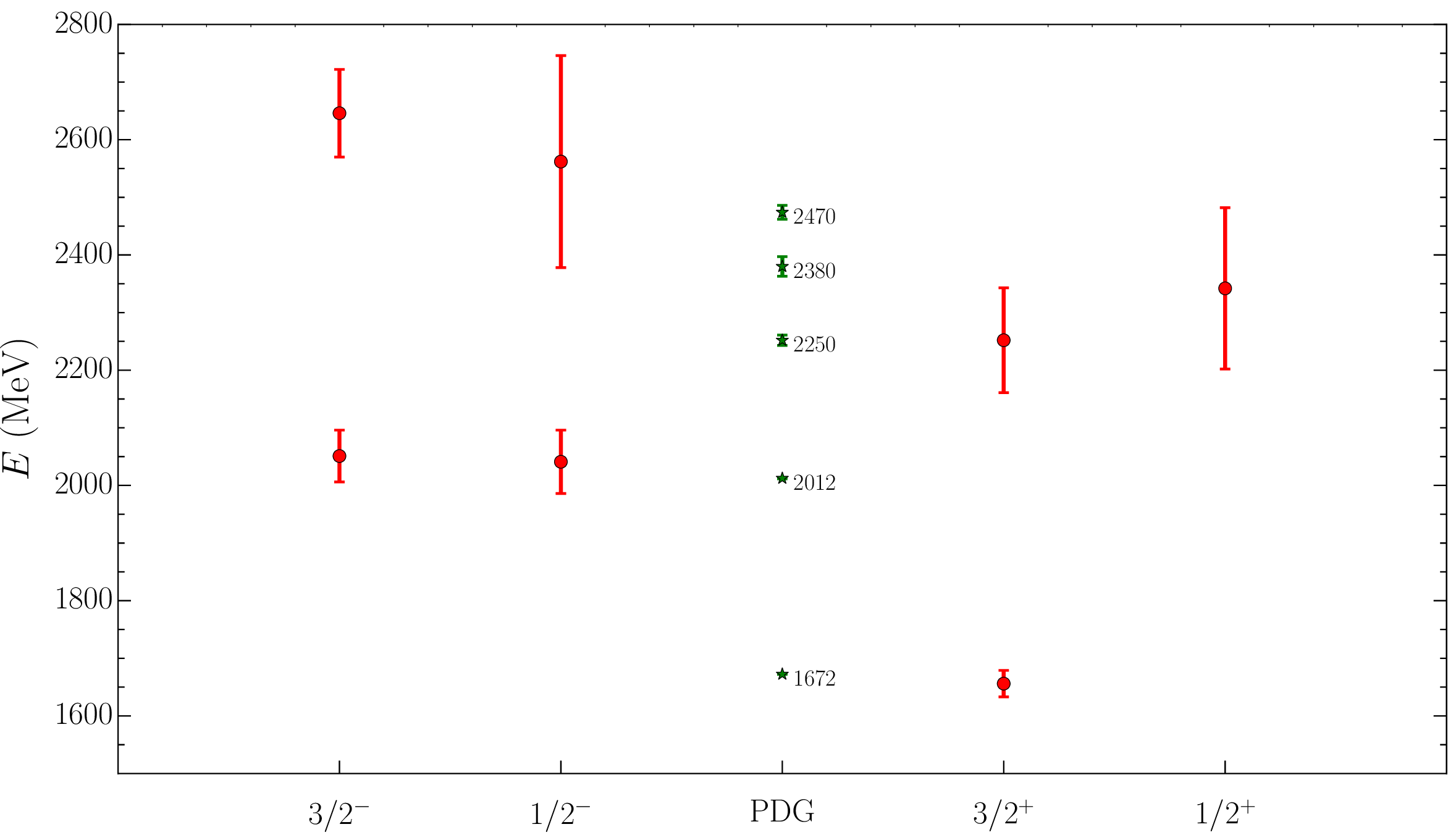}
		\caption{Plot of our extrapolated masses for the low-lying $ \Omega $-baryon spectrum (red circles). Each column contains results for the listed pairings of $ J^P $ quantum numbers, with odd- and even-parity results placed on opposite sides of the PDG resonances (green stars).}
		\label{fig:pdg_comparison}
	\end{figure}

	While the first few resonances in the spectrum have narrow widths on the order of $ 10 $ MeV, the others have widths of about 50 MeV or more. In a complete calculation, the greater number of thresholds available for various processes at higher energies should be taken into account, for example in a multi-hadron interpolator approach incorporating $ \Xi^* K $ channels in order to account for shifts in the finite-volume energy levels due to nearby scattering states. 
	
	In this exploratory work, we have seen good evidence with regard to various quantum number assignments, and identification of the radial excitations is a new contribution. Notably, in both the $ 3/2^+ $ and $ 3/2^- $ spectra where we have best signal, there is an approximately equal energy splitting between radial excitations of about $ 600-700 $ MeV. With an additional energy gap of about $ 400 $ MeV between successive $ 3/2^+ $ and $ 3/2^- $ states, this gives some indication of an underlying harmonic oscillator model for three quarks. Much like the nucleon and $ \Delta $ spectra discussed in Refs.~\cite{Hockley:2023yzn}, this suggests a properly tuned constituent quark model may give useful insight into the spectrum of $ \Omega $ baryons.
	
	\section{Conclusion} \label{sec:Conclusion}
	We have presented a new investigation into the $ \Omega $-baryon spectrum relying on first principles calculations in lattice QCD. The experimentally observed resonances have relatively narrow widths, and while a full calculation involving multi-particle states is necessary for resolving details of avoided level crossings in the spectrum, an exploratory analysis such as ours offers new insight into this regime. Motivated by the scarcity of experimental information concerning the $ \Omega $ baryons, we have made several predictions for the unknown quantum numbers in the spectrum.
	
	Using a variational analysis approach based on smeared 3-quark interpolators, we have presented the first few energy levels across the even- and odd-parity $ \Omega $ sectors for both the spin-1/2 and spin-3/2 cases. We found a spin-3/2 positive parity state which was in agreement with both the experimentally measured value of the $ \Omega^-(1672) $ and with contemporary lattice QCD results.
	
	We then investigated the odd-parity spectrum, finding that the $ \Omega^-(2012) $ resonance can be associated with two odd-parity states of spin-1/2 and spin-3/2. This suggests that there are two overlapping resonances being reported in the PDG as the single $ \Omega^-(2012) $. 
	
	For the resonances above 2 GeV, we find evidence that the $ \Omega^-(2250) $ is an even-parity resonance, with a $ J^P $ assignment favouring $ 3/2^+ $. Further investigation with a broader basis of interpolating fields targeting the spin-1/2 sector may be required for disentangling the spin states. The $ \Omega^-(2380) $ and $ \Omega^-(2470) $ favour the assignment of $ J=1/2 $, particularly for the $ \Omega^-(2470) $.
	
	Aside from improved lattice QCD techniques, our understanding of the $ \Omega $-baryon spectrum will benefit from new experimental searches at facilities like J-PARC which are able to produce the data needed for constraining theoretical models. In order to resolve the nature of the $ \Omega^-(1672) $ and $ \Omega^-(2012) $ resonances in L\"{u}scher-like methods, experimental insight into the phase shifts in scattering of $ \Xi K $ and $ \Xi \pi K $ multiparticle states is required. In light of this, we find ourselves in the position of being able to make predictions of future experimental searches. However it is through formalisms such as Hamiltonian Effective Field Theory and L\"uscher's method that one can truly connect lattice QCD results with experimental data. With the advent of several next-generation colliders approaching, the resonance physics in the $ \Omega $-baryon spectrum may then be confronted with both theory and experiment on equal footing.
	
	
	\ack 
	This work was supported by an Australian Government Research Training Program Scholarship and with supercomputing resources provided by the Phoenix HPC service at the University of Adelaide. This research was undertaken with the assistance of resources from the National Computational Infrastructure (NCI), provided through the National Computational Merit Allocation Scheme, and supported by the Australian Government through Grant No. LE190100021 via the University of Adelaide Partner Share. This research was supported by the Australian Research Council through Grants No. DP190102215 and DP210103706 (DBL), and DP230101791 (AWT). WK was supported by the Pawsey Supercomputing Centre through the Pawsey Centre for Extreme Scale Readiness (PaCER) program.
	
	
	\appendix
	

	\section*{References}
	\bibliographystyle{unsrturl}
	\bibliography{bibliography}
	
\end{document}